\documentclass[11pt]{article}
\usepackage[margin=1.1in]{geometry}
\usepackage{amsmath,amsthm}
\usepackage{mathtools}
\usepackage{amssymb}
\usepackage{graphicx}
\usepackage[colorlinks=true, allcolors=blue]{hyperref}
\usepackage{natbib}
\usepackage{subfig}
\usepackage{multirow}
\usepackage{multicol}
\usepackage{tikz}
\usetikzlibrary{arrows,shapes.arrows,shapes.geometric,shapes.multipart,decorations.pathmorphing,positioning,shapes.swigs,}
\title{\bf Bounds for causal mediation effects}

\usepackage{authblk}
\author[1]{Marie S. Breum\thanks{corresponding author: masb@sund.ku.dk}}
\author[2,3]{Vanessa Didelez}
\author[1,4]{Erin E. Gabriel}
\author[1,4]{Michael C. Sachs}
\affil[1]{Section of Biostatistics, University of Copenhagen, Copenhagen, Denmark}
\affil[2]{Leibniz Institute for Prevention Research and Epidemiology – BIPS,  Bremen, Germany}
\affil[3]{Faculty of Mathematics and Computer Science, University of Bremen, Bremen, Germany}
\affil[4]{The Pioneer Centre for SMARTbiomed, University of Copenhagen, Copenhagen, Denmark}

\newcommand\independent{\protect\mathpalette{\protect\independenT}{\perp}}
\def\independenT#1#2{\mathrel{\rlap{$#1#2$}\mkern2mu{#1#2}}}

\begin{document}

\maketitle

\begin{abstract}
Several frameworks have been proposed for studying causal mediation analysis. What these frameworks have in common is that they all make assumptions for point identifications that can be violated even when treatment is randomized. When a causal effect is not point-identified, one can sometimes derive bounds, i.e. a range of possible values that are consistent with the observed data. In this work, we study causal bounds for mediation effects under both the natural effects framework and the separable effects framework. 
In particular, we show that when there are unmeasured confounders for the intermediate variables(s) the sharp symbolic bounds on separable (in)direct effect coincide with existing bounds for natural (in)direct effects in the analogous setting. We compare these bounds to valid bounds for the natural direct effects when only the cross-world independence assumption does not hold. Furthermore, we demonstrate the use and compare the results of the bounds on data from a trial investigating the effect of peanut consumption on the development of peanut allergy in infants through specific pathways of measured immunological biomarkers. 
\end{abstract}


\section{Introduction} \label{sec:introduction}
  
  The natural (in)direct effect framework proposed by \cite{robins1992identifiability} and \cite{pearl2001direct} defines direct and indirect effects in terms of cross-world counterfactual outcomes, i.e., counterfactual outcomes in a world where the exposure is assigned to a particular value and the mediator is assigned to its counterfactual value under a possibly different treatment assignment. The natural direct effect (NDE) is defined as the average effect on the outcome of setting the exposure to the active value versus the reference value when the mediator is set to what it would have been under the reference value of the exposure. The natural indirect effect (NIE) is the average effect on the outcome if the exposure was fixed at the active value but the mediator was set to what it would have been if the exposure had been set to the active value versus if exposure had been set to the reference value. A property of natural (in)direct effects is that the NDE and NIE provide a decomposition of the total treatment effect (TE) without any parametric assumptions. However, the identification of natural (in)direct effects relies on the so-called cross-world independence assumption, which has been criticized for being overly restrictive in many settings, but mainly for being impossible to empirically verify even experimentally. The cross-world independence assumption states that the counterfactual outcome under an intervention that assigns the exposure to a particular value is independent of the counterfactual mediator values under a conflicting intervention on the exposure (possibly conditionally on measured covariates). This assumption is violated in the presence of measured or unmeasured mediator-outcome confounders if these are affected by the exposure \citep{avin2005identifiability}, which limits the practical applicability of natural (in)direct effects. The cross-world assumption can also be violated in other ways than through intermediate confounding \citep{robins2010alternative,AndrewsDidelez2021}. 
  
  Without the cross-world assumption natural (in)direct effect are not nonparametrically identified \citep{robins1992identifiability}. \cite{robins2010alternative} derived symbolic bounds for the NDE under violations of the cross-world assumption, in the absence of an intermediate confounder. \cite{tchetgen2014bounds}  provide bounds on the NDE in the presence of a measured post-treatment confounder and \cite{miles2017partial} generalize the bounds of \cite{tchetgen2014bounds} to settings with a polytomous mediator. In this work, we show that the bounds of \cite{robins2010alternative} are valid, but not necessarily sharp, as they are identical to bounds based on the  Fr\'echet bounds. Here `sharp' means that the bounds are attainable. `Valid' means that no value outside the bounds is a possible value given the true probabilities that are estimable from the data.  The definition of natural (in)direct effects has been extended to the setting with two sequential or independent mediators \citep{daniel2015causal}. We also derive valid bounds on the NDE for two sequential mediators when only the cross-world assumption is violated. To our knowledge, these bounds have not been previously considered in the literature. 

  Another framework for causal mediation analysis is the interventionist framework of \cite{robins2010alternative, RobinsRichardsonShpitser2020interventionist}, which focuses on different estimands without reference to cross-world counterfactuals. In particular, this approach assumes that the exposure has two binary components: a ‘direct’ one that affects the outcome of interest not through the mediator and an ‘indirect’ one that only affects the outcome of interest through its effect on the mediator. By assuming that the two components can be intervened upon separately, they define interventionist analogs of natural direct and indirect effects. Among other, this has important implications for the longitudinal and survival outcome case \citep{didelez2019defining}. Interventionist (in)direct effects are also commonly referred to as separable (in)direct effects, a term coined by \cite{Stensrud2020separable, Stensrud2021separable} in a series of papers on causal estimands in a competing event setting. Unlike the natural direct effect, identification of the separable/interventionist (in)direct effects requires only `single world independence' assumptions (in the hypothetical expanded world where the treatment can be decomposed into the two components). The assumptions are, at least in principle, empirically verifiable in a future four-arm trial where the treatment components are assigned different values in addition to being identical \citep{RobinsRichardsonShpitser2020interventionist}. In the case of an intermediate confounder, the definition of the separable effects estimand depends on whether the intermediate confounder is directly affected by the ‘direct’ component, the ‘indirect’ component, or something else, such as a third component \citep{robins2010alternative}. 

   If, as is often the case, there are unmeasured confounders for the intermediate variables and the outcome, then neither the natural nor the separable (in)direct effects are point identified.
   \citet{sjolander2009bounds} derived symbolic bounds on the natural direct effect for a single mediator when only the treatment randomization assumption holds. \cite{sachs2023general} developed a general approach for deriving symbolic bounds using linear programming techniques and characterized a class of settings in which the method is guaranteed to provide sharp valid bounds. The method was implemented in the causaloptim R-package \citep{jonzon2023accessible}. The bounds of \cite{sjolander2009bounds} satisfy the conditions  necessary for the bounds to be valid and sharp. \cite{Gabriel23} used the method of \cite{sachs2023general} to derive bounds for the natural (in)direct effects for two sequential or independent mediators. 
   In this work, we use the method of \cite{sachs2023general} to derive sharp valid bounds on separable effects, assuming only treatment randomization, and we show that these coincide with existing bounds for natural (in)direct effects when only assuming randomization. 
   We also compare these bounds to valid bounds for the natural direct effects when only the cross-world independence assumption is violated, a comparison that has not previously been made.

   The manuscript is organized as follows. In Section \ref{sec:preliminaries} we review the difference between single-world and cross-world causal models and introduce some set-up and notation. In Section \ref{sec:estimands} we define the separable effect estimands and discuss their identification and relation to natural (in)direct effects. In Section \ref{sec:cross-world}, we comment on the use of the cross-world assumption in the derivation of previous bounds. In Section \ref{sec:bounds}, we present our novel bounds results and in Section \ref{sec:relation} we discuss the relation between our novel bounds and previous bounds. In Section \ref{sec:dataexample}, we illustrate the use of the bounds on data from a trial investigating the effect of peanut consumption on the development of peanut allergy in infants. Some discussion and final remarks are provided in Section \ref{sec:discussion}.

\section{Preliminaries} \label{sec:preliminaries}

\subsection{`Single world' versus `cross worlds' causal models}

\begin{figure}[ht]
\begin{center}
        \begin{tikzpicture}
        \tikzset{line width=1.5pt, outer sep=0pt, ell/.style={draw,fill=white, inner sep=2pt, line width=1.5pt}, swig vsplit={gap=5pt, inner line width right=0.5pt}};
			\node[name=A, ell, shape=ellipse]{$A$};
                \node[name=M, ell, shape=ellipse, right=10mm of A]{$M$};
                \node[name=Y, ell, shape=ellipse, below=10mm of M]{$Y$};
      \draw[->,line width=0.5pt,>=stealth] (A) to (M);
            \draw[->,line width=0.5pt,>=stealth] (A) to (Y);
            \draw[->,line width=0.5pt,>=stealth] (M) to (Y);

              \end{tikzpicture}
\end{center}
    \caption{DAG with node set $\boldsymbol{V}=(A, M, Y)$.}
    \label{fig:DAG}
\end{figure}
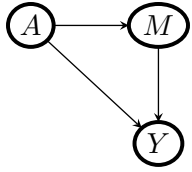

Let $\boldsymbol{V}$ be a set of random variables and $\mathcal{G}(\boldsymbol{V})$ a directed acyclic graph (DAG) involving those variables. As an example, consider the DAG in Figure \ref{fig:DAG} with node set $\boldsymbol{V}=(A, M, Y)$. Moreover, we let $\text{pa}_\mathcal{G}(V)$ be the set of parents of $V \in \boldsymbol{V}$ in $\mathcal{G}(\boldsymbol{V})$. 
For example, in Figure \ref{fig:DAG}, $\text{pa}_\mathcal{G}(Y) = \{A, M\}$, $\text{pa}_\mathcal{G}(M) = \{A\}$ and $\text{pa}_\mathcal{G}(A) = \emptyset$. Furthermore, let $\boldsymbol{x}_{\boldsymbol{R}}$ denote the subset of $\boldsymbol{x} \in \text{supp}(\boldsymbol{V})$ corresponding to the subset $\boldsymbol{R} \subseteq \boldsymbol{V}$. In particular, $\boldsymbol{x}_{pa_\mathcal{G}(V)}$ is the subset of $\boldsymbol{x}$ corresponding to the set of parents of $V \in \boldsymbol{V}$.

We let $V(\text{\textbf{pa}}_V)$ be the counterfactual value of $V$ under an intervention assigning the value(s) of $\text{pa}_\mathcal{G}(V)$ to $\text{\textbf{pa}}_V$. For example, in Figure \ref{fig:DAG}, $M(a)$ is the counterfactual value of $M$ under an intervention that sets $A$ to the value $a$ and $Y(a,m)$ is the counterfactual value of $Y$ under the intervention that sets $A$ to $a$ and $M$ to $m$.
More generally, for any $V \in \boldsymbol{V}$ and $\boldsymbol{R} \subseteq \boldsymbol{V}$, let $V(\boldsymbol{r})$ be the value of $V$ if we had, possibly contrary to the fact, set $\boldsymbol{R}$ to the value(s) $\boldsymbol{r}$ by intervention.  When $\boldsymbol{R} \neq \text{pa}_\mathcal{G}(V)$, then $V(\boldsymbol{r})$ is defined recursively via $V(\boldsymbol{r})=V\left(\boldsymbol{r}_{\text{pa}_\mathcal{G}(V) \cap \boldsymbol{R}}, (\text{\textbf{PA}}_V \backslash \boldsymbol{R})(\boldsymbol{r}) \right)$ where $ (\text{\textbf{PA}}_V \backslash \boldsymbol{R})(\boldsymbol{r}) = \left\{V^*(\boldsymbol{r}) \mid V^* \in \text{pa}_\mathcal{G}(V), V^* \notin \boldsymbol{R} \right\}$. For example, in Figure \ref{fig:DAG}, the counterfactual value of $Y$ under an intervention that sets $A$ to the value $a$, $Y(a)$, is defined as $ Y(a, M(a))$.

A causal model is a set of distributions over $\{V(\boldsymbol{x}_{pa_\mathcal{G}(V)}) \mid V \in \boldsymbol{V} \}$ defined by some restrictions. We focus on two types of causal models: the Nonparametric Structural Equation Models with Independent Errors (NPSEM-IE) of \cite{Pearl2000book} and the Finest Fully Randomized Causally Interpretable Structured Tree Graph Model (FFRCISTGM) \citep{robins1986new, robins2010alternative}.

The NPSEM-IE is defined by satisfying the restriction that the sets $\{V(\boldsymbol{x}_{pa_\mathcal{G}(V)}) \mid  \forall \ \boldsymbol{x}_{pa_\mathcal{G}(V)} \}$ are mutually independent for all $V \in \boldsymbol{V}$. In particular, the NPSEM-IE model associated with the DAG in Figure \ref{fig:DAG} would impose the following. 
\begin{align}
    \label{eq:NPSEM-IE}
    A \independent M(a') \independent Y(a, m) \text{ for each set of values } a, a', m.
\end{align}
In particular, the NPSEM-IE implies the cross-world independence assumption $Y(a,m) \independent M(a')$ which states that the counterfactual outcome in the world where the exposure is set to the level $a$ is independent of the counterfactual mediator value in the world where the exposure is set to the level $a'$. This is why the NPSEM-IE is sometimes called a `cross worlds' model. Another way to view the NPSEM-IE is to consider the nonparametric structural equations associated with the DAG in Figure \ref{fig:DAG}:
\begin{equation}
    \label{eq:NPSEM}
    A=f_A(\epsilon_A), \quad M=f_M(A, \epsilon_M), \quad Y=f_Y(A, M, \epsilon_Y),
\end{equation}
where $\epsilon=(\epsilon_A, \epsilon_M, \epsilon_Y)$ are exogenous variables and $f=(f_A, f_M, f_Y)$ are deterministic functions. The potential outcomes are obtained as $M(a)=f_M(a, \epsilon_M)$ and $Y(a,m)=f_Y(a,m, \epsilon_Y)$ and therefore the nested counterfactuals are obtained as $Y(a,M(a'))=f_Y(a,f_M(a', \epsilon_M), \epsilon_Y)$. The NPSEM-IE restriction in \eqref{eq:NPSEM-IE} is then equivalent to the assumption that the exogenous variables are mutually independent
\begin{align*}
    \epsilon_A \independent \epsilon_M \independent \epsilon_Y.
\end{align*}

 In contrast, the FFRCISTG model is defined by satisfying the restriction that the sets $\{V(\boldsymbol{r}) \mid V \in \boldsymbol{V}, \boldsymbol{r} = \boldsymbol{x}_{pa_\mathcal{G}(V)}\}$ are mutually independent for all $\boldsymbol{x} \in \text{supp}(\boldsymbol{V})$. The FFRCISTG model associated with the DAG in Figure \ref{fig:DAG} would impose the following. 
\begin{align}
    \label{eq:FFRCISTG}
    A \independent M(a) \independent Y(a, m) \text{ for each set of values } a,  m.
\end{align}

The FFRCISTG model is a `single world' model in the sense that it only imposes restrictions that are (in principle) possible to verify, and \eqref{eq:FFRCISTG} is refered to as the single world no confounding assumption \citep{shpitser2022multivariate}. Note that the NPSEM-IE is a submodel of the FFRCISTG model, since the assumptions in \eqref{eq:NPSEM-IE} contain the assumptions in \eqref{eq:FFRCISTG}. The FFRCISTGM can also be formulated via the nonparametric structural equations in \eqref{eq:NPSEM}, but with weaker conditions on the exogenous variables. 

   \subsection{Setting I: simple mediation setting}
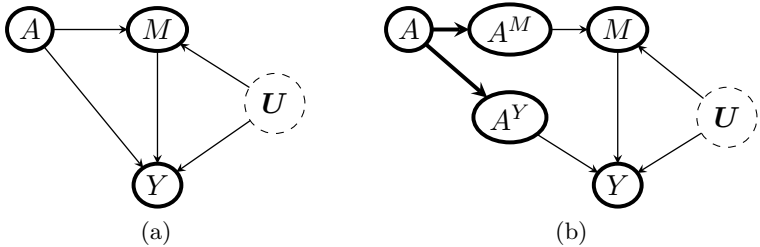
\begin{figure}[ht]
	\begin{center}
        \subfloat[]{
        \begin{tikzpicture}
        \tikzset{line width=1.5pt, outer sep=0pt, ell/.style={draw,fill=white, inner sep=2pt, line width=1.5pt}, swig vsplit={gap=5pt, inner line width right=0.5pt}};
			\node[name=A, ell, shape=ellipse]{$A$};
                \node[name=M, ell, shape=ellipse, right=10mm of A]{$M$};
                \node[name=Y, ell, shape=ellipse, below=15mm of M]{$Y$};
                \node[name=U, draw,dashed,circle,below right =5mm and 10mm of M]{$\boldsymbol{U}$};
      \draw[->,line width=0.5pt,>=stealth] (A) to (M);
            \draw[->,line width=0.5pt,>=stealth] (A) to (Y);
            \draw[->,line width=0.5pt,>=stealth] (M) to (Y);
             \draw[->,line width=0.5pt,>=stealth] (U) to (Y);
              \draw[->,line width=0.5pt,>=stealth] (U) to (M);
              \end{tikzpicture}
  \label{fig:setIa}
  }
  \qquad
   \subfloat[]{
		\begin{tikzpicture}
        \tikzset{line width=1.5pt, outer sep=0pt, ell/.style={draw,fill=white, inner sep=2pt, line width=1.5pt}, swig vsplit={gap=5pt, inner line width right=0.5pt}};
			\node[name=A, ell, shape=ellipse]{$A$};
                \node[name=AM, ell, shape=ellipse, right=5mm of A]{$A^M$};
                \node[name=AY, ell, shape=ellipse, below=5mm of AM]{$A^Y$};
                \node[name=M, ell, shape=ellipse, right=5mm of AM]{$M$};
                \node[name=Y, ell, shape=ellipse, below=15mm of M]{$Y$};
                \node[name=U, draw,dashed,circle, right=20mm of AY]{$\boldsymbol{U}$};
      \draw[->,line width=1.5pt,>=stealth] (A) to (AM);
      \draw[->,line width=1.5pt,>=stealth] (A) to (AY);
      \draw[->,line width=0.5pt,>=stealth] (AM) to (M);
            \draw[->,line width=0.5pt,>=stealth] (AY) to (Y);
            \draw[->,line width=0.5pt,>=stealth] (M) to (Y);
             \draw[->,line width=0.5pt,>=stealth] (U) to (Y);
              \draw[->,line width=0.5pt,>=stealth] (U) to (M);

		\end{tikzpicture}
  \label{fig:setIb}
  }
	\end{center}
	\caption{(a) Simple DAG. (b) expanded DAG. Thick arrows indicate deterministic relations.}
        \label{fig:setI}
    \end{figure}

   Let $A$ be a binary exposure of interest and $Y$ be a binary outcome of interest. We let $M$ be a binary intermediate variable and $\boldsymbol{U}$ be a set of unmeasured confounders that affect $M$ and $Y$. This is illustrated in the DAG in Figure \ref{fig:setIa}. We assume that the exposure $A$ can be separated into two binary components which we will denote $A^M$ and $A^Y$, where the component $A^M$ only affects the outcome of interest through its effect on the intermediate variable $M$, and the component $A^Y$ only affects the outcome of interest directly. This corresponds to the absence of edges from $A^M$ to $Y$ and from $A^Y$ to $M$ in the expanded graph in Figure \ref{fig:setIb}. In the observed data, the two components are deterministically related $A^M=A^Y=A$. This is illustrated by the thick arrows in Figure \ref{fig:setIb}. 
   
 The nonparametric structural equation model (NPSEM) associated with the DAG in Figure \ref{fig:setIb} is
\begin{equation}
   \begin{gathered}
       A=F_A(\epsilon_A), \quad A^M \equiv A, \quad A^Y \equiv A,\quad M=F_M(A^M, \boldsymbol{U}, \epsilon_M), \\
       Y = F_Y(A^Y, M, \boldsymbol{U}, \epsilon_Y),
   \end{gathered}    
   \label{eq:SCM1}
   \end{equation}
   where $\epsilon= (\epsilon_A, \epsilon_M, \epsilon_Y)$ and $U$ are exogenous variables and $F= (F_A, F_M, F_Y)$ are deterministic functions. Because we are not assuming anything about $U$, this model allows for violations of cross-world independence relations that would be implied by the NPSEM-IE if $U$ were observed. 


    \subsection{Setting II: post-treatment confounding}
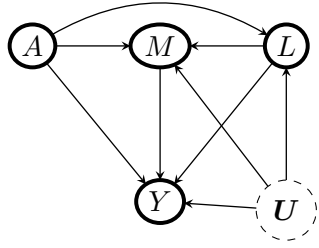
\begin{figure}[ht]
    \centering
    \begin{tikzpicture}
    \tikzset{line width=1.5pt, outer sep=0pt, ell/.style={draw,fill=white, inner sep=2pt, line width=1.5pt}, swig vsplit={gap=5pt, inner line width right=0.5pt}};
    \node[name=A, ell, shape=ellipse]{$A$};
    \node[name=M, ell, shape=ellipse,  right=10mm of A]{$M$};
    \node[name=L, ell, shape=ellipse,  right=10mm of M]{$L$};
    \node[name=Y, ell, shape=ellipse, below=15mm of M]{$Y$};
    \node[name=U, draw,dashed,circle, below=15mm of L]{$\boldsymbol{U}$}; 
    \draw[->,line width=0.5pt,>=stealth] (A) to (M);
    \draw[->,line width=0.5pt,>=stealth] (A) to (Y);
    \draw[->,line width=0.5pt,>=stealth] (M) to (Y);
    \draw[->,line width=0.5pt,>=stealth] (U) to (Y);
    \draw[->,line width=0.5pt,>=stealth] (U) to (M);
    \draw[->,line width=0.5pt,>=stealth] (L) to (M);
    \draw[->,line width=0.5pt,>=stealth] (L) to (Y);
    \draw[->,line width=0.5pt,>=stealth] (U) to (L);
    \draw[->,line width=0.5pt,>=stealth, bend left] (A) to (L);
    \end{tikzpicture}
    \caption{Simple DAG.}
    \label{fig:setII}
    \end{figure}

    We will also consider a different setting where $A, M$ and $Y$ are defined as before and where we let $L$ be an observed common cause of $M$ and $Y$ which is affected by $A$. Here $L$ may be a post-treatment confounder in the traditional sense. Alternatively, it can be seen to represent a simple version of the longitudinal mediation setting with $L$ being the first measurement of the mediator and $M$ the second, or this may represent the setting where the outcome is longitudinal with $L$ being the first and $Y$ the second measurement. We let $\boldsymbol{U}$ be a set of unmeasured confounders that affect $L$, $M$, and $Y$. This setting is illustrated in the causal graph in Figure \ref{fig:setII}. 

    The post-treatment confounder $L$ may be directly affected by either component $A^M$, $A^Y$ or a third component $A^L$:
    \begin{description}
        \item[(a)] Exposure $A$ can be separated into two components $A^M$ and $A^Y$, such that the direct effect of $A$ on $Y$ (not via $M$ and $L$) is solely attributed to $A^Y$ and the indirect effect of $A$ on $Y$ via $M$ and $L$ is solely attributed to $A^M$.
        \item[(b)] Exposure $A$ can be separated into two components $A^M$ and $A^Y$, such that the direct effect of $A$ on $Y$ and the indirect effect of $A$ on $Y$ via $L$ are solely attributed to $A^Y$ and the indirect effect of $A$ on $Y$ via $M$ (not including the path $L \to M \to Y$) is solely attributed to $A^M$.
                \item[(c)] Exposure $A$ can be separated into three different components $A^L$, $A^M$, and $A^Y$, such that the direct effect of $A$ on $Y$ (not via $M$ and $L$) is solely attributed to $A^Y$, the indirect effect of $A$ on $Y$ via $M$ (not including the path $L \to M \to Y$) is solely attributed to $A^M$, and the indirect of $A$ on $Y$ via $L$ (including the path $L \to M \to Y$) is solely attributed to $A^L$.
    \end{description}

\begin{figure}[ht]
       \centering

       \subfloat[]{
		\begin{tikzpicture}
        \tikzset{line width=1.5pt, outer sep=0pt, ell/.style={draw,fill=white, inner sep=2pt, line width=1.5pt}, swig vsplit={gap=5pt, inner line width right=0.5pt}};
			\node[name=A, ell, shape=ellipse]{$A$};
                \node[name=AM, ell, shape=ellipse, right=5mm of A]{$A^M$};
                \node[name=AY, ell, shape=ellipse, below=5mm of AM]{$A^Y$};
                \node[name=M, ell, shape=ellipse, right=5mm of AM]{$M$};
                \node[name=L, ell, shape=ellipse, right=10mm of M]{$L$};
                \node[name=Y, ell, shape=ellipse, below=15mm of M]{$Y$};
                \node[name=U, draw,dashed,circle, right=10mm of Y]{$\boldsymbol{U}$};
      \draw[->,line width=1.5pt,>=stealth] (A) to (AM);
      \draw[->,line width=1.5pt,>=stealth] (A) to (AY);
      \draw[->,line width=0.5pt,>=stealth] (AM) to (M);
            \draw[->,line width=0.5pt,>=stealth] (AY) to (Y);
            \draw[->,line width=0.5pt,>=stealth, bend left] (AM) to (L);
            \draw[->,line width=0.5pt,>=stealth] (M) to (Y);
             \draw[->,line width=0.5pt,>=stealth] (U) to (Y);
              \draw[->,line width=0.5pt,>=stealth] (U) to (M);
\draw[->,line width=0.5pt,>=stealth] (U) to (L);
\draw[->,line width=0.5pt,>=stealth] (L) to (M);
\draw[->,line width=0.5pt,>=stealth] (L) to (Y);
		\end{tikzpicture}
   \label{fig:setII_AM}
  } \quad
       \subfloat[]{
		\begin{tikzpicture}
        \tikzset{line width=1.5pt, outer sep=0pt, ell/.style={draw,fill=white, inner sep=2pt, line width=1.5pt}, swig vsplit={gap=5pt, inner line width right=0.5pt}};
			\node[name=A, ell, shape=ellipse]{$A$};
                \node[name=AM, ell, shape=ellipse, right=5mm of A]{$A^M$};
                \node[name=AY, ell, shape=ellipse, below=5mm of AM]{$A^Y$};
                \node[name=M, ell, shape=ellipse, right=5mm of AM]{$M$};
                \node[name=L, ell, shape=ellipse, right=10mm of M]{$L$};
                \node[name=Y, ell, shape=ellipse, below=15mm of M]{$Y$};
                \node[name=U, draw,dashed,circle, right=10mm of Y]{$\boldsymbol{U}$};
      \draw[->,line width=1.5pt,>=stealth] (A) to (AM);
      \draw[->,line width=1.5pt,>=stealth] (A) to (AY);
      \draw[->,line width=0.5pt,>=stealth] (AM) to (M);
            \draw[->,line width=0.5pt,>=stealth] (AY) to (Y);
            \draw[->,line width=0.5pt,>=stealth] (AY) to (L);
            \draw[->,line width=0.5pt,>=stealth] (M) to (Y);
             \draw[->,line width=0.5pt,>=stealth] (U) to (Y);
              \draw[->,line width=0.5pt,>=stealth] (U) to (M);
\draw[->,line width=0.5pt,>=stealth] (U) to (L);
\draw[->,line width=0.5pt,>=stealth] (L) to (M);
\draw[->,line width=0.5pt,>=stealth] (L) to (Y);
		\end{tikzpicture}
   \label{fig:setII_AY}
  }\\
             \subfloat{
		\begin{tikzpicture}
        \tikzset{line width=1.5pt, outer sep=0pt, ell/.style={draw,fill=white, inner sep=2pt, line width=1.5pt}, swig vsplit={gap=5pt, inner line width right=0.5pt}};
			\node[name=A, ell, shape=ellipse]{$A$};
                \node[name=AM, ell, shape=ellipse, right=5mm of A]{$A^M$};
                \node[name=AY, ell, shape=ellipse, below=5mm of AM]{$A^Y$};
                \node[name=AL, ell, shape=ellipse, above=5mm of AM]{$A^L$};
                \node[name=M, ell, shape=ellipse, right=5mm of AM]{$M$};
                \node[name=L, ell, shape=ellipse, right=10mm of M]{$L$};
                \node[name=Y, ell, shape=ellipse, below=15mm of M]{$Y$};
                \node[name=U, draw,dashed,circle, right=10mm of Y]{$\boldsymbol{U}$};
      \draw[->,line width=1.5pt,>=stealth] (A) to (AM);
      \draw[->,line width=1.5pt,>=stealth] (A) to (AY);
      \draw[->,line width=1.5pt,>=stealth] (A) to (AL);
      \draw[->,line width=0.5pt,>=stealth] (AM) to (M);
            \draw[->,line width=0.5pt,>=stealth] (AY) to (Y);
            \draw[->,line width=0.5pt,>=stealth] (AL) to (L);
            \draw[->,line width=0.5pt,>=stealth] (M) to (Y);
             \draw[->,line width=0.5pt,>=stealth] (U) to (Y);
              \draw[->,line width=0.5pt,>=stealth] (U) to (M);
\draw[->,line width=0.5pt,>=stealth] (U) to (L);
\draw[->,line width=0.5pt,>=stealth] (L) to (M);
\draw[->,line width=0.5pt,>=stealth] (L) to (Y);
		\end{tikzpicture}
   \label{fig:setII_AL}
  }
  \caption{(a) Expanded DAG where $L$ is directly affected by $A^M$. (b) Expanded DAG where $L$ is directly affected by $A^Y$. (a) Expanded DAG where $L$ is directly affected by $A^L$. Thick arrows indicate deterministic relations.}
   \end{figure}
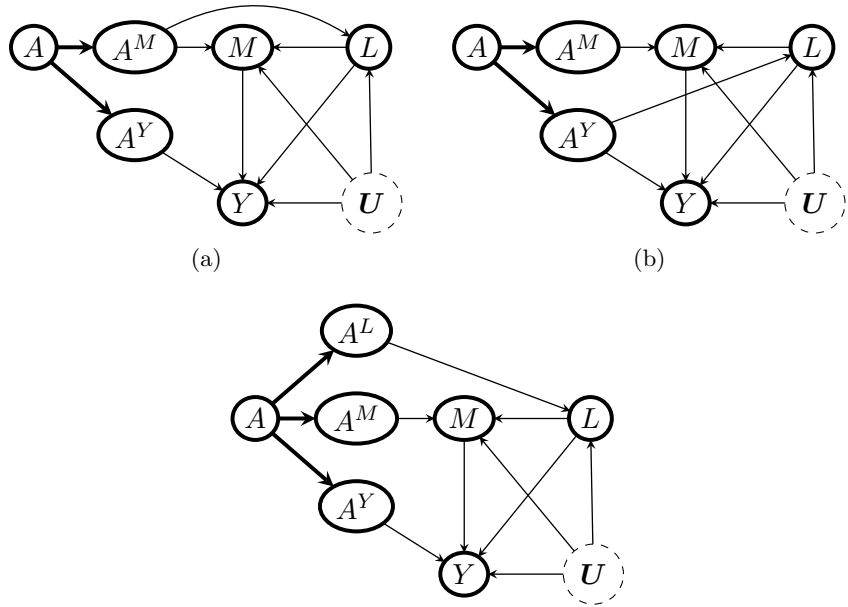

    The different scenarios are illustrated in the expanded graphs in Figures \ref{fig:setII_AM}-\ref{fig:setII_AL}.
    We note that the scenarios are special cases of the isolation conditions considered by \cite{Stensrud2021separable}, who develop separable effect estimands in a competing event setting with time-varying confounders.
        
    The NPSEM associated with the DAG in Figure \ref{fig:setII_AL} is
   \begin{equation}
   \begin{gathered}
       A=F_A(\epsilon_A), \quad  A^L \equiv A,  \quad A^M \equiv A, A^Y \equiv A, \quad L=F_L(A^L, \boldsymbol{U}, \epsilon_L), \\
       M=F_M(A^M, L, \boldsymbol{U}, \epsilon_M), \quad  Y = F_Y(A^Y, M, L, \boldsymbol{U},  \epsilon_Y),
   \end{gathered}    
   \label{eq:SCM2}
   \end{equation}
   where $\epsilon= (\epsilon_A, \epsilon_L, \epsilon_M, \epsilon_Y)$ and $U$ are exogenous variables and $F=(F_A, F_L, F_M, F_Y)$ are deterministic functions. 

   The NPSEM associated with Figure \ref{fig:setII_AM} is obtained by replacing $A^L$ with $A^M$ in \eqref{eq:SCM2} and the NPSEM associated with Figure \ref{fig:setII_AY} is obtained by replacing $A^L$ with $A^Y$ in \eqref{eq:SCM2}.

\section{Estimand(s) and identification} \label{sec:estimands}

    \subsection{Setting I}   

    Consider an intervention to set $A^Y=a$ and $A^M=a'$ in the NPSEM in \eqref{eq:SCM1}. The associated counterfactuals $M(A^M=a')$ and $Y(A^Y=a, A^M=a')$ are:
   \begin{align*}
   \begin{split}
       M(A^M=a')&=F_M(a', \boldsymbol{U}, \epsilon_M), \\
       Y(A^Y=a, A^M=a') &= F_Y(a, M(a'), \boldsymbol{U}, \epsilon_Y).
   \end{split}    
   \end{align*}

    Then, the separable direct effects (SDE) are defined as  
    \begin{align}
    \label{eq:SDE_I}
        \mbox{SDE}(a)= Pr\left\{Y(A^Y=1, A^M=a)=1\right\} - Pr\left\{Y(A^Y=0, A^M=a)=1\right\},
    \end{align}
    and the separable indirect effects (SIE) as
        \begin{align}
    \label{eq:SIE_I}
        \mbox{SIE}(a)= Pr\left\{Y(A^Y=a, A^M=1)=1\right\} - Pr\left\{Y(A^Y=a, A^M=0)=1\right\},
    \end{align}
    for $a \in \{0, 1\}$. 

    Note that the NPSEM in \eqref{eq:SCM1} implies that the hypothetical treatment intervention is such that $Y(A^Y=a, A^M=a) = Y(A=a)$. This in turn implies that the SDE and SIE provide a decomposition of the total treatment effect (TE) as $\mbox{TE}=\mbox{SDE}(0)+\mbox{SIE}(1) = \mbox{SDE}(1)+\mbox{SIE}(0)$.

    Let $M(A=a), Y(A=a, M=m)$ be the counterfactuals associated with the DAG in Figure \ref{fig:setIa} under a NPSEM-IE. As shown in \cite{robins2010alternative} \eqref{eq:SDE_I} can be aligned to the counterfactuals associated with Figure \ref{fig:setIa} as follows.
    \begin{align}
    \label{eq:NDE}
        Pr\left[Y\left\{A=1, M(A=a)\right\} = 1 \right] - Pr\left[Y\left\{A=0, M(A=a)\right\} = 1 \right],
    \end{align}
    for $a \in \{0, 1\}$. This is the definition of the natural direct effect, which we will denote $\mbox{NDE}(a)$. 

   Similarly \eqref{eq:SIE_I} can be aligned to the counterfactuals associated with Figure \ref{fig:setIa} as
    \begin{align}
    \label{eq:NIE}
        Pr\left[Y\left\{A=a, M(A=1)\right\} = 1 \right] - Pr\left[Y\left\{A=a, M(A=0)\right\} = 1 \right],
    \end{align}
    for $a \in \{0, 1\}$. This is the definition of the natural indirect effect, which we will denote $\mbox{NIE}(a)$. 


   \subsubsection{Identifying functional}
    If $\boldsymbol{U}$ is measured, then the SDE and SIE can be identified from the data under the following assumptions.
    \begin{description}
        \item[A0.I)] Treatment randomization: (i) $M(a) \independent A$, \quad (ii) $Y(a, m) \independent A$.
        \item[A1.I)] Dismissible components: (i) $M(a^M) \independent A^Y \mid A^M=a^M, \boldsymbol{U}$, \quad (ii) $Y(a^M, a^Y) \independent A^M \mid M=m, A^Y=a^Y, \boldsymbol{U}$.
    \end{description}

\begin{figure}[ht]
      \centering
      \begin{tikzpicture}
      \tikzset{line width=1.5pt, outer sep=0pt, ell/.style={draw,fill=white, inner sep=2pt, line width=1.5pt}, swig vsplit={gap=5pt, inner line width right=0.5pt}};
      \node[name=A, ell, shape=ellipse]{$A$};
      \node[name=aM,shape=swig vsplit, right=3mm of A]{
      \nodepart{left}{$A^M$}
      \nodepart{right}{$a^M$} };
      \node[name=aY,shape=swig vsplit, below=4mm of aM]{
      \nodepart{left}{$A^Y$}
      \nodepart{right}{$a^Y$} };
      \node[name=M, ell, shape=ellipse,  right=4mm of aM]{$M(a^M)$};
      \node[name=Y, ell, shape=ellipse,  below=10mm of M]{$Y(a^M, a^Y)$};
      \draw[->,line width=1.5pt,>=stealth] (A) to (aM);
      \draw[->,line width=1.5pt,>=stealth] (A) to (aY);
      \draw[->, line width=0.5pt,>=stealth] (aM) to (M);
     \draw[->, line width=0.5pt,>=stealth] (aY) to (Y);
     \draw[->, line width=0.5pt,>=stealth] (M) to (Y);
     \node[name=U,draw,dashed,circle, right=35mm of aY]{$\boldsymbol{U}$};
     \draw[->, line width=0.5pt,>=stealth] (U) to (M.east);
     \draw[->, line width=0.5pt,>=stealth] (U) to (Y.east);
\end{tikzpicture}
\caption{SWIG corresponding to an intervention that sets $A^M$ to $a^M$ and $A^Y$ to $a^Y$ in \ref{fig:setIb}. Thick arrows indicate deterministic relations.}
        \label{fig:swigI}
  \end{figure}
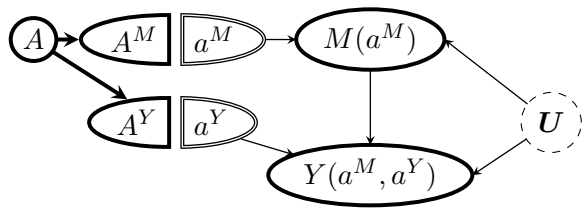

    These assumptions are implied by the FFRCISTG model associated with the NPSEM in \eqref{eq:SCM1}, and can be read off the single world intervention graph (SWIG) \citep{richardson2013single} in Figure \ref{fig:swigI}. 
In particular, if $\boldsymbol{U}$ is measured, then under  A0.I)-A1.I) we have
    \begin{align*}
         &Pr\left\{Y(A^Y=a, A^M=a')=1 \right\}\\
        =&\sum_{m, \boldsymbol{u}} Pr\left\{Y=1 \mid A=a, M=m, \boldsymbol{U}=\boldsymbol{u} \right\}  Pr \left\{ M=m \mid A=a', \boldsymbol{U}=\boldsymbol{u} \right\} Pr\left\{\boldsymbol{U}=\boldsymbol{u}\right\}, 
    \end{align*}
    from which the identifying functionals of \eqref{eq:SDE_I} and \eqref{eq:SIE_I} follow. This identification result was first shown in \cite{RobinsRichardson2011}. For completeness, we give a proof in the Supplementary Materials. 
    This is the same identifying functional as for the nested counterfactual $Pr \left\{Y(a, M(a'))=1\right\}$ under a different set of identifying assumptions implied by the NPSEM-IE associated with the DAG in Figure \ref{fig:setIa}. 

    When $\boldsymbol{U}$ is unmeasured the SDE and SIE are not point identified because versions of the dismissible components conditions in A1.I) that do not condition on $\boldsymbol{U}$ do not hold.  

    \subsection{Setting II}
    The definition of natural (in)direct effects has been extended to the setting with two sequential or independent mediators. \cite{daniel2015causal} defines the natural direct effects for two sequential mediators, with $L$ being the first and $M$ the second, as
    \begin{multline*}
        \mbox{NDE}(a',a'',a''') \\ =Pr\left\{Y(1, L(a'), M(a'', L(a''')))=1 \right\} -Pr\left\{Y(0, L(a'), M(a'', L(a''')))=1 \right\},
    \end{multline*}
    for $a', a'', a''' \in \{0,1\}$.

    Moreover, they define the indirect effect through $L$ only as
    \begin{multline*}
        \mbox{NIE}_L(a, a'', a''') \\ =Pr\left\{Y(a, L(1), M(a'', L(a''')))=1 \right\} -Pr\left\{Y(a, L(0), M(a'', L(a''')))=1 \right\},
    \end{multline*}
    the indirect effect through $M$ only as
     \begin{multline*}
        \mbox{NIE}_M(a, a', a''') \\ =Pr\left\{Y(a, L(a'), M(1, L(a''')))=1 \right\} -Pr\left\{Y(a, L(a'), M(0, L(a''')))=1 \right\},
    \end{multline*}
    and the indirect effect through both $M$ and $L$ as
    \begin{multline*}
        \mbox{NIE}_{LM}(a, a', a'') \\ =Pr\left\{Y(a, L(a'), M(a'', L(1)))=1 \right\} -Pr\left\{Y(a, L(a'), M(a'', L(0)))=1 \right\},
    \end{multline*}
    for $a, a', a'', a''' \in \{0,1\}$.

    In the following, we define the separable effect estimands in the extended models (a)-(c) and discuss how they relate to the natural (in)direct effects above. 
   
    \subsubsection{Expanded model (a)}

    Consider now an intervention to set $A^Y=a$ and $A^M=a'$ in the NPSEM associated with the DAG in Figure \ref{fig:setII_AM}. The associated counterfactuals  $L(A^M=a')$, $ M(A^M=a')$, $Y(A^Y=a, A^M=a')$ are 
   \begin{align*}
        L(A^M=a') &= F_L(a', \epsilon_L), \\
        M(A^M=a') &= F_M(a', L(A^M=a') , \boldsymbol{U}, \epsilon_M), \\
        Y(A^Y=a, A^M=a') &= F_Y(a,  M(A^M=a'), L(A^M=a'), \boldsymbol{U}, \epsilon_Y). 
    \end{align*}
    The SDE can be defined as in \eqref{eq:SDE_I} as the contrast $\mbox{SDE}(a)= Pr\left\{Y(A^Y=1, A^M=a)=1\right\} - Pr\left\{Y(A^Y=0, A^M=a)=1\right\}$ and can be aligned to the NDE in terms of the counterfactuals associated with Figure \ref{fig:setII} under the NPSEM-IE as
    \begin{align}
    \label{eq:NDE_IIa}
     Pr\left\{Y(1, L(a), M(a, L(a)))=1 \right\} -Pr\left\{Y(0, L(a), M(a, L(a)))=1 \right\},
   \end{align}
 for $a \in \{0,1\}$. Note that this is the definition of the natural direct effects for two sequential mediators NDE$(a, a, a)$, as given in \cite{daniel2015causal}. 

 The SIE can be defined as in \eqref{eq:SIE_I} as the contrast $\mbox{SIE}(a)= Pr\left\{Y(A^Y=a, A^M=1)=1\right\} - Pr\left\{Y(A^Y=a, A^M=0)=1\right\}$ and can be aligned to the NIE in terms of the counterfactuals associated with Figure \ref{fig:setII} under the NPSEM-IE as
    \begin{align}
    \label{eq:NIE_IIa}
     Pr\left\{Y(a, L(1), M(1, L(1)))=1 \right\} -Pr\left\{Y(a, L(0), M(0, L(0)))=1 \right\},
   \end{align}
   for $a \in \{0,1\}$. This indirect effect captures the effect not through the direct path from $A$ to $Y$,  and can be decomposed as, e.g. $\mbox{NIE}_L(a, 1, 1)$ + $\mbox{NIE}_M(a, 0, 1)$ + $\mbox{NIE}_{LM}(a, 0, 0)$.

   \paragraph{Identifying functional}
   If $\boldsymbol{U}$ is measured, the identifying functional for the counterfactual $Pr\left\{Y(A^Y=a, A^M=a')=1 \right\}$ is 
    \begin{align*}
        &Pr\left\{Y(A^Y=a, A^M=a')=1 \right\} \\ =&\sum_{m, \ell, \boldsymbol{u}} Pr\left\{Y=1 \mid A=a, M=m, L=\ell, \boldsymbol{U}=\boldsymbol{u} \right\} Pr \left\{ M=m \mid A=a', L=\ell, \boldsymbol{U}=\boldsymbol{u} \right\} \\
        &\times Pr \left\{ L=\ell \mid A=a', \boldsymbol{U}=\boldsymbol{u} \right\}Pr\left\{\boldsymbol{U}=\boldsymbol{u}\right\},
    \end{align*}
    from which the identifying functionals for the SDEs and SIEs defined above follow. Identification follows by the following independencies that can be read off the SWIG in Figure \ref{fig:swigII_AM}:
    \begin{description}
        \item[A0.II)] Treatment randomization: (i) $L(a) \independent A$, \quad (ii) $M(a, \ell) \independent A$, \quad (iii) $Y(a, \ell, m) \independent A$.
        \item[A1.IIa)] Dismissible components: (i) $L(a^M) \independent  A^Y \mid A^M=a^M, \boldsymbol{U}$, \quad (ii) $M(a^M) \independent A^Y \mid L=\ell, A^M=a^M, \boldsymbol{U}$, \quad (iii) $Y(a^M, a^Y) \independent A^M \mid L=\ell, M=m, A^Y=a^Y, \boldsymbol{U}$.
        \end{description}

\sloppy The above identifying functional is identical to the one for the nested counterfactual $Pr \left\{Y(a, L(a'), M(a', L(a')))=1\right\}$ under a different set of identifying assumptions implied by the NPSEM-IE associated with Figure \ref{fig:setII}. 

    \subsubsection{Expanded model (b)}
    Similarly, the counterfactuals $L(A^Y=a)$, $M(A^Y=a, A^M=a')$, $Y(A^Y=a, A^M=a')$ obtained by setting $A^Y=a$ and $A^M=a'$ in the NPSEM associated with the DAG in Figure \ref{fig:setII_AY} are  
    \begin{align*}
        L(A^Y=a) &= F_L(a, \epsilon_L), \\
        M(A^Y=a, A^M=a') &= F_M(a', L(A^Y=a), \boldsymbol{U}, \epsilon_M), \\
        Y(A^Y=a, A^M=a') &= F_Y(a,  M(A^Y=a, A^M=a'), L(A^Y=a), \boldsymbol{U}, \epsilon_Y).
    \end{align*}
    The SDE can be defined as in \eqref{eq:SDE_I} as the contrast $\mbox{SDE}(a)= Pr\left\{Y(A^Y=1, A^M=a)=1\right\} - Pr\left\{Y(A^Y=0, A^M=a)=1\right\}$ and can be aligned to the NDE in terms of the counterfactuals associated with Figure \ref{fig:setII} under the NPSEM-IE as follows.
    \begin{align}
    \label{eq:NDE_IIb}
     Pr\left\{Y(1, L(1), M(a, L(1)))=1 \right\} -Pr\left\{Y(0, L(0), M(a, L(0)))=1 \right\},
   \end{align}
    for $a \in \{0,1\}$. This direct effect captures the effect not through $M$ and can be decomposed into as e.g. $\mbox{NDE}(1, a, 1) + \mbox{NIE}_L(0, a, 1) + \mbox{NIE}_{LM}(0, 0, a)$.

    The SIE can be defined as in \eqref{eq:SIE_I} as the contrast $\mbox{SIE}(a)= Pr\left\{Y(A^Y=a, A^M=1)=1\right\} - Pr\left\{Y(A^Y=a, A^M=0)=1\right\}$ and can be aligned to the NIE in terms of the counterfactuals associated with Figure \ref{fig:setII} under the NPSEM-IE as follows.
    \begin{align}
    \label{eq:NIE_IIb}
     Pr\left\{Y(a, L(a), M(1, L(a)))=1 \right\} -Pr\left\{Y(a, L(a), M(0, L(a)))=1 \right\},
   \end{align}
    for $a \in \{0,1\}$. This is the definition of the indirect effect through $M$ given in \cite{daniel2015causal} $\mbox{NIE}_M(a, a, a)$. 

    Note that when $Y$ is t-year survival and $L$ is past survival, then the definition of the SDEs and SIEs in the expanded model (b) corresponds to the definitions in \cite{didelez2019defining}.

       \paragraph{Identifying functional}
   If $\boldsymbol{U}$ is measured then the identifying functional for the counterfactual $Pr\left\{Y(A^Y=a, A^M=a')=1 \right\}$ is 
    \begin{align*}
        &Pr\left\{Y(A^Y=a, A^M=a')=1 \right\} \\ =&\sum_{m, \ell, \boldsymbol{u}} Pr\left\{Y=1 \mid A=a, M=m, L=\ell, \boldsymbol{U}=\boldsymbol{u} \right\} Pr \left\{ M=m \mid A=a', L=\ell, \boldsymbol{U}=\boldsymbol{u} \right\} \\
        &\times Pr \left\{ L=\ell \mid A=a, \boldsymbol{U}=\boldsymbol{u} \right\}Pr\left\{\boldsymbol{U}=\boldsymbol{u}\right\},
    \end{align*}
        from which the identifying functionals for the SDEs and SIEs defined above follow. Indentification follows by the treatment randomization assumption (Assumption A0.II)), and the following independencies that can be read off the SWIG in Figure \ref{fig:swigII_AY} :
        \begin{description}
        \item[A1.IIb)] Dismissible components: (i)  $L(a^Y) \independent  A^M \mid A^Y=a^Y, \boldsymbol{U}$, \quad (ii) $M(a^M) \independent A^Y \mid L=\ell, A^M=a^M, \boldsymbol{U}$, \quad (iii) $Y(a^M, a^Y) \independent A^M \mid L=\ell, M=m, A^Y=a^Y, \boldsymbol{U}$.
    \end{description}

\sloppy The above identifying functional is identical to the one for the nested counterfactual $Pr \left\{Y(a, L(a), M(a', L(a)))=1\right\}$ under a different set of identifying assumptions implied by the NPSEM-IE associated with Figure \ref{fig:setII}.

    \subsubsection{Expanded model (c)}
    Next,  consider an intervention to set $A^Y=a$, $A^M=a'$ and $A^L=a''$ in the NPSEM that corresponds to the setting in Figure \ref{fig:setII_AL}. The associated counterfactuals $L(A^L=a'')$, $M(A^M=a', A^L=a'')$ and $Y(A^Y=a, A^M=a', A^L=a'')$ are
    \begin{align*}
    L(A^L=a'')&=F_L(a'', \epsilon_L), \\
    M(A^M=a', A^L=a'')&=F_M(a', F_L(a'', \epsilon_L), \boldsymbol{U}, \epsilon_M),\\
    Y(A^Y=a, A^M=a', A^L=a'')&=F_Y(a, M(A^M=a', A^L=a''), L(A^L=a''), \boldsymbol{U}, \epsilon_Y),
    \end{align*}

    We can then define SDEs via the contrast:
    \begin{multline}
            \label{eq:SDE_II}
        SDE(a', a'', a''')\\=Pr\left\{Y(A^Y=1, A^M=a', A^L=a'')=1\right\} - Pr\left\{Y(A^Y=0,  A^M=a', A^L=a''')=1\right\},
    \end{multline}
    for $a', a'', a''' \in \{0,1\}$. When $L$ is binary, as we assume in this work, the estimand in \eqref{eq:SDE_II} always reduces to the estimand in (a) or (b). In particular, when $a''=a'''$ it reduces to the estimand  in (a) and when $a'' \neq a'''$ it reduces to the estimand in (b). In more general cases where $L$ is polytomous, the estimand in \eqref{eq:SDE_II} reduces to the estimand in (a) only when $a'=a''=a'''$ and to the estimand in (b) only when $a''=1$ and $a'''=0$.

    We note that \eqref{eq:SDE_II} can be aligned to the counterfactuals associated with Figure \ref{fig:setII} under the NPSEM-IE as follows.
    \begin{align}
     \label{eq:NDE_IIc}
        Pr \left\{Y(1, L(a''), M(a', L(a''))) = 1 \right\} - Pr \left\{Y(0, L(a'''), M(a', L(a'''))) = 1 \right\} ,
    \end{align}
    for $a', a'', a''' \in \{0,1\}$. When $a'=a''=a'''$, then \eqref{eq:NDE_IIc} is identical to \eqref{eq:NDE_IIa}. When $a''=1$ and $a'''=0$ then \eqref{eq:NDE_IIc} is identical to \eqref{eq:NDE_IIb}.

    Similarly, we can define SIEs via the contrast:
    \begin{multline}
    \label{eq:SIE_II}
        SIE(a, a'', a''')\\=Pr\left\{Y(A^Y=a, A^M=1, A^L=a'')=1\right\} - Pr\left\{Y(A^Y=a,  A^M=0, A^L=a''')=1\right\},
    \end{multline}
    for $a, a'', a''' \in \{0,1\}$.  As above, when $L$ is binary, the estimand in \eqref{eq:SIE_II} always reduces to the estimand in (a) or (b).  In more general cases where $L$ is polytomous, the estimand in \eqref{eq:SDE_II} reduces to the estimand in (a) only when $a''=1$ and $a'''=0$ and to the estimand in (b) only when $a=a''=a'''$,

    We note that \eqref{eq:SIE_II} can be aligned to the counterfactuals associated with Figure \ref{fig:setII} under the NPSEM-IE as follows.
    \begin{align}
     \label{eq:NIE_IIc}
        Pr \left\{Y(a, L(a''), M(1, L(a''))) = 1 \right\} - Pr \left\{Y(a, L(a'''), M(0, L(a'''))) = 1 \right\} ,
    \end{align}
    for $a', a'', a''' \in \{0,1\}$. When $a=a''=a'''$, then \eqref{eq:NIE_IIc} is identical to \eqref{eq:NIE_IIb}. When $a''=1$ and $a'''=0$ then \eqref{eq:NIE_IIc} is identical to \eqref{eq:NIE_IIa}.

    Note that the NPSEM in \eqref{eq:SCM2} implies that the hypothetical treatment intervention is such that $Y(A^Y=a, A^M=a, A^L=a) = Y(A=a)$.  This implies that the SDEs and SIEs in \eqref{eq:SDE_II} and \eqref{eq:SIE_II} provide a decomposition of the total treatment effect.

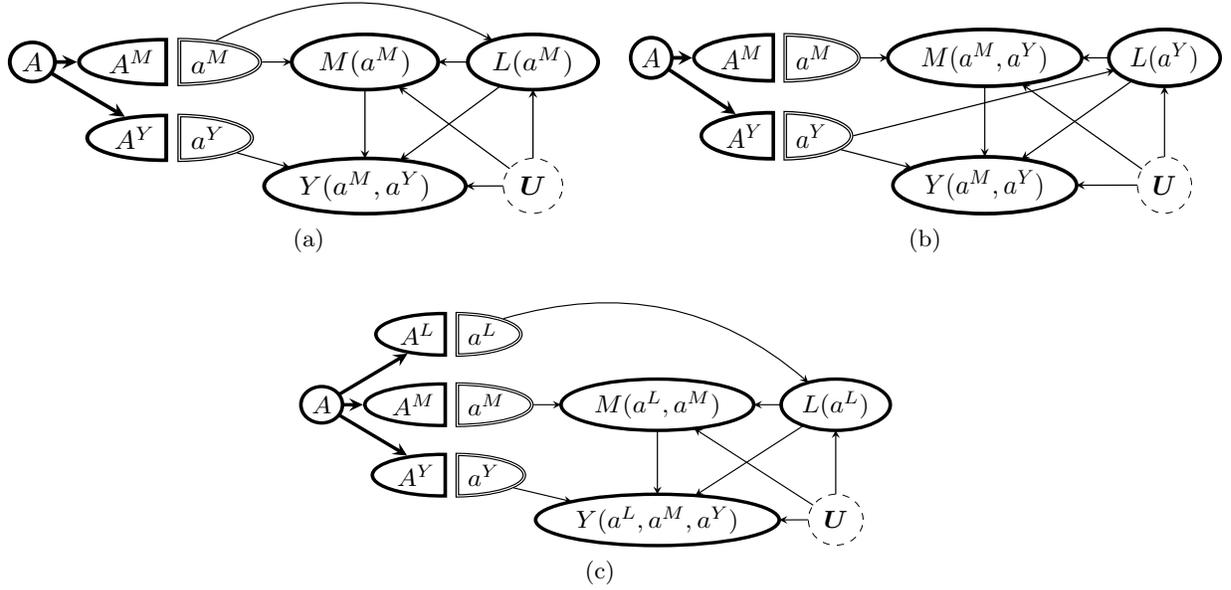
\begin{figure}
      \centering 
     
\subfloat[]{
      \resizebox{8cm}{3cm}{%
      \begin{tikzpicture}
      \tikzset{line width=1.5pt, outer sep=0pt, ell/.style={draw,fill=white, inner sep=2pt, line width=1.5pt}, swig vsplit={gap=5pt, inner line width right=0.5pt}};
      \node[name=A, ell, shape=ellipse]{$A$};
      \node[name=aM,shape=swig vsplit, right=3mm of A]{
      \nodepart{left}{$A^M$}
      \nodepart{right}{$a^M$} };
      \node[name=aY,shape=swig vsplit, below=4mm of aM]{
      \nodepart{left}{$A^Y$}
      \nodepart{right}{$a^Y$} };
          \node[name=M, ell, shape=ellipse,  right=4mm of aM]{$M(a^M)$};
      \node[name=Y, ell, shape=ellipse,  below=10mm of M]{$Y(a^M, a^Y)$};
      \node[name=L, ell, shape=ellipse,  right=4mm of M]{$L(a^M)$};
      \draw[->,line width=1.5pt,>=stealth] (A) to (aM);
      \draw[->,line width=1.5pt,>=stealth] (A) to (aY);
      \draw[->, line width=0.5pt,>=stealth] (aM) to (M);
     \draw[->, line width=0.5pt,>=stealth] (aY) to (Y);
      \draw[->, line width=0.5pt,>=stealth, bend left] (aM) to (L);
     \draw[->, line width=0.5pt,>=stealth] (M) to (Y);
     \node[name=U,draw,dashed,circle, below=10mm of L]{$\boldsymbol{U}$};
     \draw[->, line width=0.5pt,>=stealth] (U) to (M);
     \draw[->, line width=0.5pt,>=stealth] (U) to (Y.east);
     \draw[->, line width=0.5pt,>=stealth] (U) to (L);
      \draw[->, line width=0.5pt,>=stealth] (L) to (M);
       \draw[->, line width=0.5pt,>=stealth] (L) to (Y);
\end{tikzpicture}
\label{fig:swigII_AM}
}
}
\subfloat[]{
\resizebox{8cm}{2.5cm}{%
      \begin{tikzpicture}
      \tikzset{line width=1.5pt, outer sep=0pt, ell/.style={draw,fill=white, inner sep=2pt, line width=1.5pt}, swig vsplit={gap=5pt, inner line width right=0.5pt}};
      \node[name=A, ell, shape=ellipse]{$A$};
      \node[name=aM,shape=swig vsplit, right=3mm of A]{
      \nodepart{left}{$A^M$}
      \nodepart{right}{$a^M$} };
      \node[name=aY,shape=swig vsplit, below=4mm of aM]{
      \nodepart{left}{$A^Y$}
      \nodepart{right}{$a^Y$} };
          \node[name=M, ell, shape=ellipse,  right=4mm of aM]{$M(a^M, a^Y)$};
      \node[name=Y, ell, shape=ellipse,  below=10mm of M]{$Y(a^M, a^Y)$};
      \node[name=L, ell, shape=ellipse,  right=4mm of M]{$L(a^Y)$};
      \draw[->,line width=1.5pt,>=stealth] (A) to (aM);
      \draw[->,line width=1.5pt,>=stealth] (A) to (aY);
      \draw[->, line width=0.5pt,>=stealth] (aM) to (M);
     \draw[->, line width=0.5pt,>=stealth] (aY) to (Y);
      \draw[->, line width=0.5pt,>=stealth] (aY.east) to (L);
     \draw[->, line width=0.5pt,>=stealth] (M) to (Y);
     \node[name=U,draw,dashed,circle, below=10mm of L]{$\boldsymbol{U}$};
     \draw[->, line width=0.5pt,>=stealth] (U) to (M);
     \draw[->, line width=0.5pt,>=stealth] (U) to (Y.east);
     \draw[->, line width=0.5pt,>=stealth] (U) to (L);
      \draw[->, line width=0.5pt,>=stealth] (L) to (M);
       \draw[->, line width=0.5pt,>=stealth] (L) to (Y);
\end{tikzpicture}
\label{fig:swigII_AY}
}}\\
      \subfloat[]{
      \resizebox{8cm}{3.5cm}{%
      \begin{tikzpicture}
      \tikzset{line width=1.5pt, outer sep=0pt, ell/.style={draw,fill=white, inner sep=2pt, line width=1.5pt}, swig vsplit={gap=5pt, inner line width right=0.5pt}};
      \node[name=A, ell, shape=ellipse]{$A$};
      \node[name=aM,shape=swig vsplit, right=3mm of A]{
      \nodepart{left}{$A^M$}
      \nodepart{right}{$a^M$} };
      \node[name=aY,shape=swig vsplit, below=4mm of aM]{
      \nodepart{left}{$A^Y$}
      \nodepart{right}{$a^Y$} };
     \node[name=aL,shape=swig vsplit, above=4mm of aM]{
      \nodepart{left}{$A^L$}
      \nodepart{right}{$a^L$} };
          \node[name=M, ell, shape=ellipse,  right=4mm of aM]{$M(a^L, a^M)$};
      \node[name=Y, ell, shape=ellipse,  below=10mm of M]{$Y(a^L, a^M, a^Y)$};
      \node[name=L, ell, shape=ellipse,  right=4mm of M]{$L(a^L)$};
      \draw[->,line width=1.5pt,>=stealth] (A) to (aM);
      \draw[->,line width=1.5pt,>=stealth] (A) to (aY);
      \draw[->,line width=1.5pt,>=stealth] (A) to (aL);
      \draw[->, line width=0.5pt,>=stealth] (aM) to (M);
     \draw[->, line width=0.5pt,>=stealth] (aY) to (Y);
      \draw[->, line width=0.5pt,>=stealth, bend left] (aL) to (L);
     \draw[->, line width=0.5pt,>=stealth] (M) to (Y);
     \node[name=U,draw,dashed,circle, below=10mm of L]{$\boldsymbol{U}$};
     \draw[->, line width=0.5pt,>=stealth] (U) to (M);
     \draw[->, line width=0.5pt,>=stealth] (U) to (Y.east);
     \draw[->, line width=0.5pt,>=stealth] (U) to (L);
      \draw[->, line width=0.5pt,>=stealth] (L) to (M);
       \draw[->, line width=0.5pt,>=stealth] (L) to (Y);
\end{tikzpicture}
\label{fig:swigII_AL}
}
} 

\caption{(a) SWIG corresponding to an intervention that sets $A^M$ to $a^M$ and $A^Y$ to $a^Y$ in Fig \ref{fig:setII_AM}. (b) SWIG corresponding to an intervention that sets $A^M$ to $a^M$ and $A^Y$ to $a^Y$ in Fig \ref{fig:setII_AY}. (c) SWIG corresponding to an intervention that sets $A^M$ to $a^M$, $A^Y$ to $a^Y$ and $A^L$ to $a^L$ in Fig \ref{fig:setII_AL}. Thick arrows indicate deterministic relations.}
        \label{fig:swigII}
  \end{figure}

\paragraph{Identifying functional}
If $\boldsymbol{U}$ is observed, then the SDEs and SIEs defined above are identified under the assumptions that can be read off the SWIGs in Figure \ref{fig:swigII_AL}.  In particular, the SWIG in Figure \ref{fig:swigII_AL}, in addition to the treatment randomization assumption A0.II), encodes the following conditional independencies,
\begin{description}
        \item[A1.IIc)] Dismissible components: (i) $L(a^L, a^M, a^Y) \independent (A^M, A^Y) \mid A^L=a^L, \boldsymbol{U}$, \quad (ii) $M(a^L, a^M, a^Y) \independent (A^L, A^Y) \mid L=\ell, A^M=a^M, \boldsymbol{U}$, \quad (iii) $Y(a^L, a^M, a^Y) \independent (A^L, A^M) \mid L=\ell, M=m, A^Y=a^Y, \boldsymbol{U}$.
    \end{description}
If $\boldsymbol{U}$ is measured, then under treatment randomization and A1.IIc) the identifying functional for the counterfactual $Pr\left\{Y(A^Y=a, A^M=a', A^L=a'')=1 \right\}$ is 
    \begin{align*}
        &Pr\left\{Y(A^Y=a, A^M=a', A^L=a'')=1 \right\} \\ =&\sum_{m, \ell, \boldsymbol{u}} Pr\left\{Y=1 \mid A=a, M=m, L=\ell, \boldsymbol{U}=\boldsymbol{u} \right\} Pr \left\{ M=m \mid A=a', L=\ell, \boldsymbol{U}=\boldsymbol{u} \right\} \\
        &\times Pr \left\{ L=\ell \mid A=a'', \boldsymbol{U}=\boldsymbol{u} \right\}Pr\left\{\boldsymbol{U}=\boldsymbol{u}\right\},
    \end{align*}
   from which the identifying functionals for the SDEs and SIEs defined above follow. 

   \sloppy When $a=a''$ the above is the same identifying functional as for the nested counterfactual $Pr \left\{Y(a, L(a), M(a', L(a)))=1\right\}$ under a different set of identifying assumptions.  When $a'=a''$ this is the same identifying functional  as for the nested counterfactual $Pr \left\{Y(a, L(a'), M(a', L(a')))=1\right\}$ under a different set of identifying assumptions implied by the NPSEM-IE associated with Figure \ref{fig:setII}.

\section{Cross-world independencies and previous bounds for NDE and NIE} \label{sec:cross-world}

As discussed in Section \ref{sec:preliminaries}, when we do not make any assumptions about $\mbox{U}$, i.e., when $\mbox{U}$ is a general confounder, then the NPSEM-IE associated with \eqref{eq:SCM1} and \eqref{eq:SCM2} allows for violations of cross-world independence relations, that is, although we will derive bounds via the NPSEM-IE and linear program, we are not imposing the cross-world independence assumption.

\subsection{Setting I}
Previous bounds for NDEs in setting I in \citet{sjolander2009bounds} have not imposed cross-world independence of this nature, but they have additionally violated what is referred to as the single world independence assumption (SWI), i.e. $Y(a,m) \independent M(a)| A=a$. 

Within particular settings, bounds for the NDE have previously been derived under violations of the cross-world independence assumption alone, i.e., when all other conditions needed for identification hold. \citet{RobinsRichardson2011} derived valid but potentially not sharp bounds for the `pure direct effect' allowing for violations of cross-world dependence, in a absence of an intermediate confounder. 

\subsection{Setting II}
For setting II,  valid, but potentially not sharp, bounds for the NDE have been derived that extend the bounds of \citet{RobinsRichardson2011} to allow for an intermediate confounder \citep{tchetgen2014bounds}. To our knowledge, there have never been valid or sharp bounds derived in setting II for the path specific effects treating $L$ as an additional measured mediator where only the cross-world independence assumptions are violated, but all other assumptions for point identification hold. We derive valid bounds and present them in the following section. 
  
\section{Novel Bounds results for SDEs and NDEs} \label{sec:bounds}

We derive bounds on the SDE for the situation when $\boldsymbol{U}$ is unobserved using the \texttt{causaloptim} R-package \citep{sachs2023general, jonzon2023accessible}. Our settings satisfy the conditions necessary for the bounds to be valid and sharp, i.e. the estimand and constraints can be stated as a linear optimization problem. 

\subsection{Setting I}
We find that the bounds for the SDE under setting I, where there is only baseline unmeasured confounding of $M$ and $Y$, and where $L$ is the empty set, are identical to those for the NDE given in \citet{sjolander2009bounds} in the analogous setting. For completeness, we give these bounds in the Supplementary Material along with the bounds for the SIE/NIE.  This result is unsurprising given the correspondence between the identifying functional of the SDE and NDE.

In addition, we derive valid bounds for the NDE under setting I where there is only a violation of the cross-world independence assumption, while all other assumptions required for identification hold. We derive these bounds based on the Fr\'echet bounds \citep{frechet1935generalisation}. 
These bounds are identical to those derived in \citet{RobinsRichardson2011}, although it seems they were derived in a different manner. While the Fr\'echet inequalities are sharp, these bounds for the NDE are not necessarily sharp because they involve taking the sums of sharp bounds on the individual terms, which does not always preserve sharpness, as shown by \citet{Gabriel23}. 

An overview of the causal bounds for mediation effects under setting I is given in Table \ref{tab:bounds_settingI} below. 

\begin{table}[h]
\centering
\begingroup
\setlength{\tabcolsep}{10pt} 
\renewcommand{\arraystretch}{1.5} 
\begin{tabular}{llll}
\hline
\textbf{Bounds}  & \textbf{Assumptions}                                                   & \multicolumn{2}{c}{\textbf{Estimand}} \\ \cline{3-4} 
                 &                                                                        & \textbf{NDE}       & \textbf{SDE}       \\  \hline 
Sjölander Bounds & $\begin{rcases*}
M(a) \independent A \\
Y(a, m) \independent A
\end{rcases*}   \Rightarrow \text{treatment randomization}    $                                              & $\checkmark$       & $\checkmark$       \vspace{.5cm} \\
R\&R bounds      & \begin{tabular}[c]{@{}l@{}}$\begin{rcases*}
M(a) \independent A \\
Y(a, m) \independent A
\end{rcases*}   \Rightarrow \text{treatment randomization}    $   \\  $Y(a,m) \independent M(a) \mid A=a \} \Rightarrow \text{SWI} $ \end{tabular} & $\checkmark$       &                    \\ \hline
\end{tabular}
\caption{Overview of causal bounds for mediation effects under setting I.}
\label{tab:bounds_settingI}
\endgroup
\end{table}

\subsection{Setting II}
In setting II we find that the bounds for the SDEs are identical under  (a), (b) and (c), and that they coincide with the bounds for the natural direct effect for two sequential mediators given in \cite{Gabriel23}; please see the supplementary materials for more details.  This result is more surprising given the fact that the identifying functionals are not always the same.

In addition, we derive valid bounds for the NDE in setting II only under violation of the cross-world independence assumption, but where the cross-world independence assumption is violated by the addition of the intermediate confounder $L$. We derive these bounds based on the Fr\'echet bounds \citep{frechet1935generalisation}, the same as in setting I. We find these bounds are exactly the same as the bounds presented in \citet{Tchetgen2014} as the extension of the \citet{RobinsRichardson2011} above to the case of intermediate confounders.

We additionally derive Fr\'echet bounds for  the NDE in setting II, where $L$ is treated as a sequential mediator, as in equation \eqref{eq:NDE_IIa}, rather than a confounder, when the cross-world independencies assumption alone is violated and all other assumptions for point identification hold. First, note that for any $a, a', a'', a''' \in \{0, 1\}$
\begin{align*}
    & Pr\{Y(A = a, M(A = a', L(A = a'')), L(A=a''')) = 1\} = \\
   & \sum_{m \in \{0,1\}}\sum_{\ell \in \{0,1\}}\sum_{\ell' \in \{0, 1\}} Pr\{Y(a, m, \ell) = 1, M(a', \ell') = m, L(a'') = \ell', L(a''') = \ell\}. 
\end{align*}
Then applying the Fr\'echet bounds to an arbitrary term in the above sum we have
\begin{align*}
& \max\{0, Pr[Y(a, m, \ell) = 1] + Pr[M(a', \ell') = m] + Pr[L(a'') = \ell'] + Pr[L(a''') = \ell] - 3\} \leq \\
& Pr\{Y(a, m, \ell) = 1, M(a', \ell') = m, L(a'') = \ell', L(a''') = \ell\} \leq \\
& \min\{Pr[Y(a, m, \ell) = 1], Pr[M(a', \ell') = m], Pr[L(a'') = \ell'], Pr[L(a''') = \ell]\}.
\end{align*}
The marginal potential outcome probabilities are identified as 
\begin{align*}
    &Pr[Y(a, m, \ell) = 1] = Pr(Y = 1 | A = a, M = m, L = \ell), \\
    &Pr[M(a', \ell') = m] = Pr(M = m | A = a', L = \ell'), \\
    &Pr[L(a'') = \ell'] = Pr(L = \ell | A = a'').
\end{align*}
Substituting these expressions into the above yields the valid bounds on $Pr\{Y(A = a, M(A = a', L(A = a'')), L(A=a''')) = 1\}$ without making any cross-world independence assumptions. 

For the effect NDE(0,0,0)=$Pr\{Y(A = 1, M(A = 0, L(A = 0)), L(A=0)) = 1\} - Pr\{Y(A = 0) = 1\}$, the extension of the pure direct effect to two sequential mediators, some simplification occurs because $Pr\{L(a) = \ell, L(a) = \ell'\} = 0$ unless $\ell = \ell'$, and we get the following valid bounds: 
\small
\begin{equation}
\begin{split}
\label{newfrech}
\left\{\begin{array}{l}
 \min\{1, \max\{0, Pr[Y = 1 | A = 1, M = 0, L = 0] + Pr[M = 0 | A = 0, L = 0] + Pr[L = 0 | A = 0] - 2\} + \\ 
 \max\{0, Pr[Y = 1 | A = 1, M = 0, L = 1] + Pr[M = 0 | A = 0, L = 1] + Pr[L = 1 | A = 0] - 2\} +\\
 \max\{0, Pr[Y = 1 | A = 1, M = 1, L = 0] + Pr[M = 1 | A = 0, L = 0] + Pr[L = 0 | A = 0] - 2\} + \\
 \max\{0, Pr[Y = 1 | A = 1, M = 1, L = 1] + Pr[M = 1 | A = 0, L = 1] + Pr[L = 1 | A = 0] - 2\}\} -\\ 
 Pr[Y = 1 | A = 0], 
\end{array}\right\}\\
\leq NDE(0,0,0) \leq\\
\left\{\begin{array}{l}
 \min\{1, \min\{Pr[Y = 1 | A = 1, M = 0, L = 0], Pr[M = 0 | A = 0, L = 0], Pr[L = 0 | A = 0]\} + \\
 \min\{Pr[Y = 1 | A = 1, M = 0, L = 1], Pr[M = 0 | A = 0, L = 1], Pr[L = 1 | A = 0]\} + \\
 \min\{Pr[Y = 1 | A = 1, M = 1, L = 0], Pr[M = 1 | A = 0, L = 0], Pr[L = 0 | A = 0]\} + \\
 \min\{Pr[Y = 1 | A = 1, M = 1, L = 1], Pr[M = 1 | A = 0, L = 1], Pr[L = 1 | A = 0]\}\} - \\
 Pr[Y = 1 | A = 0]. 
\end{array} \right\}
\end{split}
\end{equation} 
\normalsize

An overview of the causal bounds for mediation effects under setting II is given in Table \ref{tab:bounds_settingII} below. 

\begin{table}[h]
\centering
\begingroup
\setlength{\tabcolsep}{10pt} 
\renewcommand{\arraystretch}{1.5} 
\begin{tabular}{llll}
\hline
\textbf{Bounds}  & \textbf{Assumptions}                                                   & \multicolumn{2}{c}{\textbf{Estimand}} \\ \cline{3-4} 
                 &                                                                        & \textbf{NDE}       & \textbf{SDE}       \\ \hline
Gabriel et al. bounds & $\begin{rcases*}
L(a) \independent A \\
M(a, \ell) \independent A \\
Y(a, \ell, m) \independent A
\end{rcases*}   \Rightarrow \text{treatment randomization}    $                                         & $\checkmark$       & $\checkmark$      \vspace{.5cm}  \\
Novel Fr\'echet bounds      & \begin{tabular}[c]{@{}l@{}}$\begin{rcases*}
L(a) \independent A \\
M(a, \ell) \independent A \\
Y(a, \ell, m) \independent A
\end{rcases*}   \Rightarrow \text{treatment randomization}    $ \\ $Y(a,\ell, m) \independent L(a), M(a) \mid A=a \} \Rightarrow \text{SWI} $  \end{tabular} & $\checkmark$       &                    \\ \hline
\end{tabular}
\caption{Overview of causal bounds for mediation effects under setting II.}
\label{tab:bounds_settingII}
\endgroup
\end{table}

\section{Relation of novel and previous bounds}\label{sec:relation}
\subsection{Setting I}

As pointed out in \citet{RobinsRichardson2011}, the NDE bounds of \citet{sjolander2009bounds} do not collapse to a point when the pure direct effect is identified, i.e. when it is equal to the controlled direct effect,
as they allow for unmeasured confounding. However, the bounds of  \citet{RobinsRichardson2011} do collapse to a point in this setting, thus, their bounds can clearly be narrower in this setting. They can sometimes be equal, for example, when the \citet{RobinsRichardson2011} lower bound equals $-Pr\{Y=1|A=0\}$, then the \citet{sjolander2009bounds} lower bound (equation (10) of that work) will be the first term, also equal to $-Pr\{Y=1|A=0\}$. We prove in Section \ref{app:NDEbounds_proof} in the Supplementary Materials that the bounds for the NDE given by \citet{sjolander2009bounds} are always at least as wide as the bounds given by \citet{RobinsRichardson2011}.

\subsection{Setting II}
It is interesting that the bounds for SDEs are identical under both (a), (b) and (c), and that they coincide with the bounds for the natural direct effects for two sequential mediators given in \cite{Gabriel23}, as they do not all have the same identifying functional. This shows that although they may have different functional forms under identification, the observable data contains the same information to limit their range.

In a setting with no confounding, we numerically compared the Fr\'echet bounds given in \ref{newfrech}, and the bounds for the NDE(0,0,0) given in \citet{Gabriel23}. We found, somewhat surprisingly, that the bounds allowing for unmeasured confounding are not always at least as wide as the Fr\'echet bounds. We found in some numeric examples (shown in Section \ref{app:NDEnumeric}) that either the lower and the upper bound of the Fr\'echet bounds given in \ref{newfrech} could be outside the lower/upper bound of the NDE(0,0,0) given in \citet{Gabriel23}. Although surprising, this does not indicate an error in either set of bounds as in the setting considered, neither set of bounds is sharp. We conjecture that the difference between the findings here and in setting I when comparing the Fr\'echet bounds and the bounds allowing for unmeasured confounding is due to the increased number of terms, the addition of which is what causes the Fr\'echet bounds to not be sharp, as each term itself is sharp. As both sets of bounds are valid, we can obtain narrower bounds by considering the maximum lower and minimum upper bounds between the two sets as suggested in \citet{gabriel2022causal}. 

\section{Real data example: peanut allergy study data} \label{sec:dataexample}
The peanut allergy trial \citep{du2015randomized} was a two-arm randomized trial that assigned 640 infants between the ages of 4 and 11 months who were at high risk of developing peanut allergy to either consume peanuts or avoid peanuts for the duration of the trial. The primary outcome was peanut allergy at 60 months of age, which was determined by an oral food challenge. We will analyze a subset of the data consisting of 530 infants in the intention-to-treat population who initially had negative results on the allergy skin-prick test. For this subset, the main analysis found that the risk of allergy was $0.137$ in the avoidance group and $0.019$ in the consumption group (absolute risk difference of $0.118$ with a 95\% confidence interval from $0.034$ to $0.203$).

\begin{figure}[ht]
\begin{center}
        \begin{tikzpicture}
        \tikzset{line width=1.5pt, outer sep=0pt, ell/.style={draw,fill=white, inner sep=2pt, line width=1.5pt}, swig vsplit={gap=5pt, inner line width right=0.5pt}};
			\node[name=A, align=center] {Peanut consumption \\ (0: consume, 1: avoid)};
                \node[name=M, align=center, right=10mm of A]{Immune modulation at 60 months \\ (0: high, 1: low)};
                \node[name=Y,align=center, below=10mm of M]{Peanut allergy at 60 months \\ (0: tolerant, 1: allergic)};
      \draw[->,line width=0.5pt,>=stealth] (A) to (M);
            \draw[->,line width=0.5pt,>=stealth] (A) to (Y);
            \draw[->,line width=0.5pt,>=stealth] (M) to (Y);

              \end{tikzpicture}
\end{center}
    \caption{Peanut allergy trial. DAG for simple mediation setting.}
    \label{fig:DAGpeanut}
\end{figure}
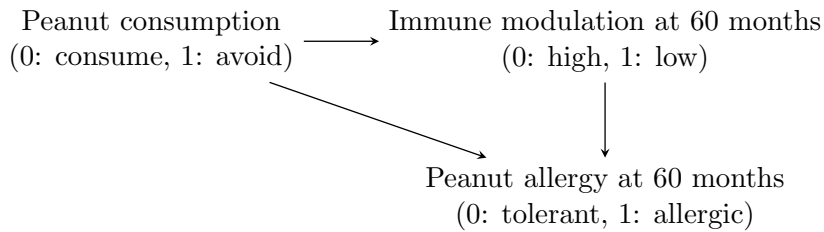

At baseline and at the scheduled follow-up visit at 12, 30, and 60 months, serum levels of peanut-specific immunoglobulin antibodies (IgE, IgG, and IgG4) were measured. The peanut-specific IgG4:IgE ratio is believed to reflect immune modulation, which is an important mediator of peanut tolerance. The assumed relationship between the variables is illustrated in Figure \ref{fig:DAGpeanut}. We want to study to what extent the effect of peanut consumption on reduction of risk of allergy is mediated through immune modulation, using the indicator of the IgG4:IgE ratio at 60 months being greater than sample mean as a mediator. An estimand that approximates this research question is the $\mbox{NDE}(0)$ which is defined as the difference in expected outcome between avoiding and consuming peanuts under an intervention that fixes immune modulation to what it would have been under peanut consumption. However, one could argue that this estimand is not well-defined because it is unclear how to set the immune modulation to a specific value (or at least there are several possible ways to do so). Suppose instead that peanut consumption can be decomposed into two components: one component $A^Y$ that affects the development of allergy directly (not through immune modulation) and one component $A^M$ that affects the development of allergy only through immune modulation. Then the $\mbox{SDE}(0)$ is the difference in expected outcome when varying the $A^Y$ component and fixing the $A^M$ component to be present. 

The point estimates of the $\mbox{SDE}(0)$ and $\mbox{NDE}(0)$ are both 0.0660, suggesting that the effect of peanut consumption on reduction of risk of allergy is not entirely mediated through high/low  IgG4:IgE ratio levels, as the trial overall indicated a positive effect on reducing allergy. However, neither the SDE nor the NDE are likely to be point identified. If we believe that only the cross-world independence assumption required to identify $\mbox{NDE}(0)$ is violated, we can use the Fréchet bounds to investigate the impact on the  $\mbox{NDE}(0)$. 
The Fréchet bounds are reported in Table \ref{tab:peanut_I}, and we see that the range is mostly positive, although it includes zero. This suggests that the effect of peanut consumption on reduction of risk of allergy is not entirely mediated through high/low  IgG4:IgE ratio levels. Note that these bounds do not allow for an ordinary confounder $U$. However, the assumption of no unmeasured mediator-outcome confounders is unlikely to hold, since both immune modulation and risk of peanut allergy are likely affected by complex environmental and biological variables. In Table \ref{tab:peanut_I} we also report bounds on the $\mbox{SDE}(0)$ and $\mbox{NDE}(0)$ assuming only treatment randomization holds. 
We see that the possible range of  $\mbox{SDE}(0)$ or $\mbox{NDE}(0)$ is both positive and negative, and covers zero. 

\begin{table}[h]
\centering
\begingroup
\setlength{\tabcolsep}{10pt} 
\renewcommand{\arraystretch}{1.5} 
\begin{tabular}{lcc}
\hline
\textbf{Assumptions}                                                     & \multicolumn{2}{c}{\textbf{Estimand}} \\ \cline{2-3} 
& \textbf{$\mbox{NDE}(0)$}       & \textbf{$\mbox{SDE}(0)$}       \\ \hline
Treatment randomization                                                  &    $[-0.322, 0.413]$                &    $[-0.322, 0.413]$                \\
\begin{tabular}[c]{@{}l@{}}Treatment randomization + \\ SWI\end{tabular} &       $[-0.0151, 0.256]$             &         $-$           \\ \hline
\end{tabular}
\caption{Peanut allergy trial. Bounds on the separable and natural direct effects in the simple mediation setting I.}
\label{tab:peanut_I}
\endgroup
\end{table}

We acknowledge that in the analysis above, we ignored earlier measurements of the IgG4:IgE ratio, which could be considered a post-treatment confounder, including the latent post-treatment confounder that we may create by dichotomizing the continuous mediators, potentially making the bounds invalid.

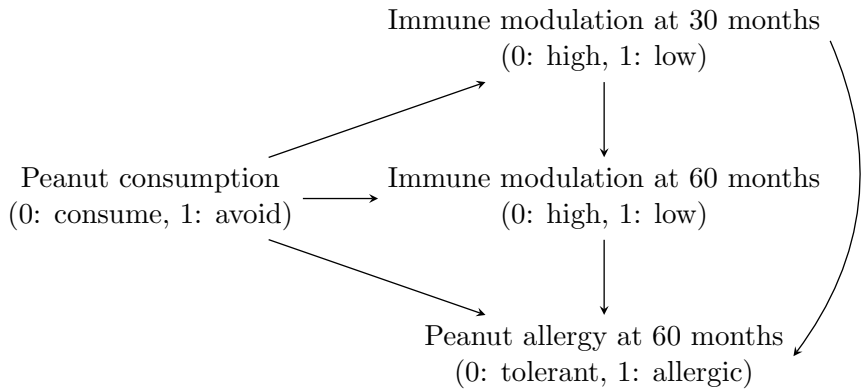
\begin{figure}[ht]
\begin{center}
        \begin{tikzpicture}
        \tikzset{line width=1.5pt, outer sep=0pt, ell/.style={draw,fill=white, inner sep=2pt, line width=1.5pt}, swig vsplit={gap=5pt, inner line width right=0.5pt}};
			\node[name=A, align=center] {Peanut consumption \\ (0: consume, 1: avoid)};
                \node[name=M, align=center, right=10mm of A]{Immune modulation at 60 months \\ (0: high, 1: low)};
                 \node[name=L, align=center, above=10mm of M]{Immune modulation at 30 months \\ (0: high, 1: low)};
                \node[name=Y,align=center, below=10mm of M]{Peanut allergy at 60 months \\ (0: tolerant, 1: allergic)};
      \draw[->,line width=0.5pt,>=stealth] (A) to (M);
            \draw[->,line width=0.5pt,>=stealth] (A) to (Y);
            \draw[->,line width=0.5pt,>=stealth] (A) to (L);
            \draw[->,line width=0.5pt,>=stealth] (M) to (Y);
            \draw[->,line width=0.5pt,>=stealth] (L) to (M);
            \draw[->,line width=0.5pt,>=stealth, bend left] (L.east) to (Y.east);

              \end{tikzpicture}
\end{center}
    \caption{Peanut allergy trial. DAG for setting II with post-treatment confounding.}
    \label{fig:DAGpeanut2}
\end{figure}

We also compute bounds where we include  the IgG4:IgE ratio at 30 months as a measured post-treatment confounder $L$. The assumed relationship between the variables is illustrated in Figure \ref{fig:DAGpeanut2}. This corresponds to the sequential mediator setting (setting II a) where the relevant estimands are $\mbox{SDE}(0,0,0)$ and $\mbox{NDE}(0,0,0)$. The point estimates are both $0.0447$.  

The Fréchet bounds for the $\mbox{NDE}(0,0,0)$ in this setting are reported in Table \ref{tab:peanut_II}. Since these are fully contained in the bounds allowing for unmeasured confounding we cannot further refine these bounds in this setting. This suggests that the range of the SDE is positive.
Although, as above, neither the SDE nor the NDE are likely to be point identified.
In Table \ref{tab:peanut_II} we also report bounds on the $\mbox{SDE}(0,0,0)$ and $\mbox{NDE}(0,0,0)$ assuming only treatment randomization holds. 
 We note that the range of the direct effect is mostly positive, although covering zero. This again, suggests that the effect of peanut consumption on reduction of risk of allergy is not entirely mediated through high/low  IgG4:IgE ratio levels.
 
 \begin{table}[h]
\centering
\begingroup
\setlength{\tabcolsep}{10pt} 
\renewcommand{\arraystretch}{1.5} 
\begin{tabular}{lcc}
\hline
\textbf{Assumptions}                                                     & \multicolumn{2}{c}{\textbf{Estimand}} \\ \cline{2-3} 
& \textbf{$\mbox{NDE}(0,0,0)$}       & \textbf{$\mbox{SDE}(0,0,0)$}       \\ \hline
Treatment randomization                                                  &    $[-0.051, 0.949] $                &    $[-0.051, 0.949] $                \\
\begin{tabular}[c]{@{}l@{}}Treatment randomization + \\ SWI\end{tabular} &       $[0, 0.277]$             &         $-$           \\ \hline
\end{tabular}
\caption{Peanut allergy trial. Bounds on the separable and natural direct effects in the setting with post-treatment confounding.}
\label{tab:peanut_II}
\endgroup
\end{table}

\section{Discussion} \label{sec:discussion}

In this work, we have studied causal bounds for mediation effects under both the natural effects framework and the separable effects framework. 
We have demonstrated that previously presented valid bounds for the NDE under violation of the cross-world independence assumption alone can be derived via Fr\'echet bounds in both our Settings I and II. This allows us to show why the bounds are valid, but not necessarily sharp. In our setting I, we demonstrate that the bounds for NDE(0) of \citet{sjolander2009bounds} are always at least as wide as the Fr\'echet bounds and thus the bounds of \citet{RobinsRichardson2011}; a comparison that has not been previously made to our knowledge. In our setting II, we derive novel Fr\'echet bounds for the path specific effect NDE(0,0,0), and show numerically that these valid but not sharp bounds are not always contained within the bounds for NDE(0,0,0) allowing for unmeasured mediators-outcome confounding \citep{gabriel2023nonparametric}. Although this is somewhat surprising, neither set of bounds is sharp and thus the refinement suggested in \citet{gabriel2022causal} of using the narrowest combined bound can be used to obtain narrower bounds in the setting with no unmeasured confounding, but violations of the cross-world independence assumption. 

Finally, we use the method of \cite{sachs2023general} to derive sharp symbolic bounds on separable (in)direct effects. We find that in the setting with a single mediator (setting I) the bounds on the separable effects are identical to the bounds on the natural effects of \cite{sjolander2009bounds}. In the setting with a post-treatment confounder (setting II) the bounds coincide with the bounds on natural effects for two sequential mediators of \cite{Gabriel23}. Although these are somewhat unsurprising results, it demonstrates that in the likely setting of unmeasured confounding, the partial identification information is the same. Thus, one should select the estimand based on interest and understandability, rather than consideration of the assumption of cross-world independence. 

As demonstrated in our real-data example, this work taken together suggests a series of estimates to present when doing mediation analysis. In the case where the desired estimand is the NDE, one would present the point estimate, a valid bound under violation of the cross-world independence assumption(s) but assuming no unmeasured confounding, and a bound for the effect allowing for unmeasured confounding. In the case where the estimand is the SDE, one would present the point estimate and a bound for the effect allowing for unmeasured confounding, which will be the same as the bound for the NDE. It is not immediately clear how one would obtain bounds for the SDE under violation of the separability assumption alone, while assuming no unmeasured confounding. This is a topic for future research. 

We have not focused on estimation; we estimate the bounds using the sample proportions, but we do not discuss inference. There has recently been an increased interest in bounds estimation. However, under the assumption that all terms in the bounds are sufficiently separated, one can use nonparametric bootstrap \citep{efron1979bootstrap} in large enough samples, as has previously been suggested for symbolic bounds \citep{gabriel2022causal, gabriel2023nonparametric}. However, in small samples and when the terms are close together, this approach may perform poorly. Other bootstrap approaches have been suggested in  \cite{ramsahai2011likelihood} and \citet{10.1093/biomet/asae020} in an attempt to correct these short comings. Instead, one may compute exact confidence intervals using the method of \cite{sachs2025improved}, which, if feasible, has been shown to provide nominal coverage.

\section*{Acknowledgments}
MSB, MCS and EEG were partially supported by a grant from Novo Nordisk fonden NNF22OC0076595 and MCS and EEG by the Pioneer Centre for SMARTbiomed. 

\section*{Data availability statement}

The trial data used in Section \ref{sec:dataexample} are publicly available and can be downloaded from the Immune Tolerance Network TrialShare website (https://www.itntrialshare.org/, study identifier: ITN032AD).

\bibliographystyle{abbrvnat}
\bibliography{bibliography}

\newpage
\centerline{\bf\LARGE Supplementary Materials}
\bigskip
\setcounter{section}{0}
\setcounter{figure}{0}
\setcounter{equation}{0}
\makeatletter
\renewcommand \thesection{S\@arabic\c@section}
\renewcommand\thetable{S\@arabic\c@table}
\renewcommand \thefigure{S\@arabic\c@figure}
\renewcommand \theequation{S\@arabic\c@equation}
\makeatother
 \setcounter{page}{1}

 \section{Identification} \label{app:id}

\subsection{Setting I}
Under setting I we have
\small
    \begin{align*}
         &Pr\left\{Y(A^Y=a^Y, A^M=a^M)=1 \right\}\\
        =&\sum_{m, \boldsymbol{u}} Pr\left\{Y(A^Y=a^Y, A^M=a^M)=1 \mid M(A^Y=a^Y, A^M=a^M)=m, \boldsymbol{U}=\boldsymbol{u} \right\} \\
        &\times Pr \left\{ M(A^Y=a^Y, A^M=a^M)=m \mid \boldsymbol{U}=\boldsymbol{u} \right\} Pr\left\{\boldsymbol{U}=\boldsymbol{u} \right\} \\
        =&\sum_{m, \boldsymbol{u}} Pr\left\{Y(A^Y=a^Y, A^M=a^Y)=1 \mid M(A^Y=a^Y, A^M=a^Y)=m, \boldsymbol{U}=\boldsymbol{u} \right\} \\
        &\times Pr \left\{ M(A^Y=a^M, A^M=a^M)=m \mid \boldsymbol{U}=\boldsymbol{u} \right\} Pr\left\{\boldsymbol{U}=\boldsymbol{u} \right\} \\
        =&\sum_{m, \boldsymbol{u}} Pr\left\{Y(A=a^Y)=1 \mid M(A=a^Y)=m, \boldsymbol{U}=\boldsymbol{u} \right\}  Pr \left\{ M(A=a^M)=m \mid \boldsymbol{U}=\boldsymbol{u} \right\} Pr\left\{\boldsymbol{U}=\boldsymbol{u} \right\} \\
          =&\sum_{m, \boldsymbol{u}} Pr\left\{Y(A=a^Y)=1 \mid A=a^Y, M=m, \boldsymbol{U}=\boldsymbol{u} \right\}  Pr \left\{ M(A=a^M)=m \mid A=a^M, \boldsymbol{U}=\boldsymbol{u} \right\} Pr\left\{\boldsymbol{U}=\boldsymbol{u}\right\} \\
        =&\sum_{m, \boldsymbol{u}} Pr\left\{Y=1 \mid A=a^Y, M=m, \boldsymbol{U}=\boldsymbol{u} \right\}  Pr \left\{ M=m \mid A=a^M, \boldsymbol{U}=\boldsymbol{u} \right\} Pr\left\{\boldsymbol{U}=\boldsymbol{u}\right\}, 
    \end{align*}
    \normalsize
where the first equality follows by the law of total probability, the second by A1.I),  the third by $Y(A^Y=a, A^M=a) = Y(A=a)$, the fourth by A0.I), and the last equality by consistency. 

\subsection{Setting II}
Under setting II (c) we have
\small
    \begin{align*}
         &Pr\left\{Y(A^Y=a^Y, A^M=a^M, A^L=a^L)=1 \right\}\\
        =&\sum_{m, \ell, \boldsymbol{u}} Pr\left\{Y(a^Y, a^M, a^L)=1 \mid L(a^Y, a^M, a^L)=\ell, M(a^Y, a^M, a^L)=m, \boldsymbol{U}=\boldsymbol{u} \right\} \\
        &\times Pr \left\{ M(a^Y, a^M, a^L)=m \mid L(a^Y, a^M, a^L)=\ell, \boldsymbol{U}=\boldsymbol{u} \right\}  Pr \left\{ L(a^Y, a^M, a^L)=m \mid \boldsymbol{U}=\boldsymbol{u} \right\}  Pr\left\{\boldsymbol{U}=\boldsymbol{u} \right\} \\
        =&\sum_{m, \ell, \boldsymbol{u}} Pr\left\{Y(a^Y, a^Y, a^Y)=1 \mid L(a^Y, a^Y, a^Y)=\ell, M(a^Y, a^Y, a^Y)=m, \boldsymbol{U}=\boldsymbol{u} \right\} \\
        &\times Pr \left\{ M(a^M, a^M, a^M)=m \mid L(a^M, a^M, a^M)=\ell, \boldsymbol{U}=\boldsymbol{u} \right\}  Pr \left\{ L(a^L, a^L, a^L)=m \mid \boldsymbol{U}=\boldsymbol{u} \right\}  Pr\left\{\boldsymbol{U}=\boldsymbol{u} \right\} \\
        =&\sum_{m, \ell. \boldsymbol{u}} Pr\left\{Y(A=a^Y)=1 \mid L(A=a^Y)=\ell, M(A=a^Y)=m, \boldsymbol{U}=\boldsymbol{u} \right\}  \\
        &\times Pr \left\{ M(A=a^M)=m \mid L(A=a^M)=\ell, \boldsymbol{U}=\boldsymbol{u} \right\} Pr\left\{L(A=a^L)=\ell \mid \boldsymbol{U}=\boldsymbol{u}  \right\}Pr\left\{\boldsymbol{U}=\boldsymbol{u} \right\} \\
          =&\sum_{m, \ell, \boldsymbol{u}} Pr\left\{Y(A=a^Y)=1 \mid A=a^Y, L=\ell, M=m, \boldsymbol{U}=\boldsymbol{u} \right\}  \\
          &\times Pr \left\{ M(A=a^M)=m \mid A=a^M, L=\ell, \boldsymbol{U}=\boldsymbol{u} \right\} Pr \left\{ L(A=a^L)=\ell, \mid A=a^L,\boldsymbol{U}=\boldsymbol{u} \right\} Pr\left\{\boldsymbol{U}=\boldsymbol{u}\right\} \\
        =&\sum_{m, \ell, \boldsymbol{u}} Pr\left\{Y=1 \mid A=a^Y, M=m, \boldsymbol{U}=\boldsymbol{u} \right\}  \\
        &\times  Pr \left\{ M=m \mid A=a^M, L=\ell, \boldsymbol{U}=\boldsymbol{u} \right\} Pr \left\{ L=\ell \mid A=a^L, \boldsymbol{U}=\boldsymbol{u} \right\}Pr\left\{\boldsymbol{U}=\boldsymbol{u}\right\}, 
    \end{align*}
    \normalsize
where the first equality follows by the law of total probability, the second by A.1.IIc), the third by A0.II), the fourth by $Y(A^Y=a, A^M=a, A^L=a) = Y(A=a)$ and the last equality by consistency. 

\section{Fréchet bounds for violation of the cross-worlds assumptions alone}

In setting I, the second term of the NDE is identified as
\[
Pr\{Y(A = 0, M(A = 0)) = 1\} = Pr\{Y = 1 | A = 0\}.
\]
The first term is 
\begin{align*}
& Pr\{Y(A = 1, M(A = 0)) = 1\} = \\ 
& Pr\{Y(A = 1, M = 0) = 1, M(A = 0) = 0\} + Pr\{Y(A = 1, M = 1) = 1, M(A = 0) = 1\}.
\end{align*}
The Fr\'echet inequality tells us that 
\begin{align*}
& \max[0, Pr\{Y(A = 1, M = m) = 1\} + Pr\{M(A = 0) = m\} - 1] \leq \\   
& Pr\{Y(A = 1, M = m) = 1, M(A = 0) = m\} \leq \\
& \min[Pr\{Y(A = 1, M = m) = 1\}, Pr\{M(A = 0) = m\}],
\end{align*}
for $m = 0, 1$. Hence 
\begin{align*}
& \max[0, Pr\{Y =1 | A = 1, M = 0\} + Pr\{M = 0 | A = 0\} - 1] + \\
& \max[0, Pr\{Y = 1 |A = 1, M = 1\} + Pr\{M = 1 | A = 0\} - 1]  - Pr\{Y = 1 | A = 0\} \leq \\   
& Pr\{Y(A = 1, M(A = 0)) = 1\} - Pr\{Y(A = 0, M(A = 0)) = 1\}\leq \\
&  \min[Pr\{Y = 1 | A = 1, M = 0\}, Pr\{M = 0 | A = 0\}] + \\
& \min[Pr\{Y = 1 | A = 1, M = 1\}, Pr\{M = 1 | A = 0\}] - Pr\{Y = 1 | A = 0\}.    
\end{align*}
These bounds are identical to those given in \citet{RobinsRichardson2011}, and presented here for completeness. 

In setting II, when $L$ is an intermediate confounder the bounds under violation of the cross-world independencies are given by:
\begin{align*}
& \max[0, Pr\{M=0|A=a*\} + \sum_{l} Pr\{Y=1|M=0, L=l, A=0\}Pr\{L=l|A=a\}-1] +\\
& \max[0, Pr\{M=0=1|A=a*\} + \sum_{l} Pr\{Y=1|M=1, L=l, A=0\}Pr\{L=l|A=a\}-1] -\\
& Pr\{Y = 1 | A = 0\} \leq \\   
& Pr\{Y(A = 1, M(A = 0)) = 1\} - Pr\{Y(A = 0, M(A = 0)) = 1\}\leq \\
&\min[Pr\{M=0|A=a*\}, \sum_{l}Pr\{Y=1|M=0,L=l,A=a\}Pr\{L=l|A=a\}] +\\
&\min[Pr\{M=1|A=a*\}, \sum_{l}Pr\{Y=1|M=1, L=l,A=a\}Pr\{L=l|A=a\}] - Pr\{Y=1|A=0\}
\end{align*}
These bounds are identical to those given in \citet{tchetgen2014identification}, 
and presented here for completeness.

\section{Bounds}\label{app:bounds}

\subsection{Setting I}
Let $p_{ym \cdot x} = Pr(Y=y, M=m \mid X=x)$. We obtained the following sharp symbolic bounds for $SDE(1)$ under Setting I, of

\begin{align}
\begin{split}
\label{eq:bounds_NDEI_1}
   \max \left\{\begin{array}{l}
-2 p_{0 0 \cdot 1}+p_{0 1 \cdot 0} -p_{0 1 \cdot 1} -p_{1 0 \cdot 1},  \\
-1+p_{0 0 \cdot 0}-p_{0 1 \cdot 1} +p_{1 0 \cdot 1}, \\
-p_{0 0 \cdot 1}-p_{0 1 \cdot 1} 
\end{array}\right\} & \leq SDE(1) \\
   &\leq \min \left\{\begin{array}{l}
p_{0 0 \cdot 0}+p_{0 1 \cdot 0} -p_{0 1 \cdot 1} +p_{1 0 \cdot 0}+p_{1 0 \cdot 1},   \\
1-p_{0 0 \cdot 1}-p_{0 1 \cdot 1},  \\
2-2p_{0 0 \cdot 1}-p_{0 1 \cdot 1} -p_{1 0 \cdot 0} -p_{1 0 \cdot 1} 
\end{array}\right\}.
\end{split}
\end{align}

These bounds are identical to those for the NDE(1) as given in \citet{sjolander2009bounds}. We obtained the following sharp symbolic bounds for $SDE(0)$ under Setting I, of 

\begin{align}
\begin{split}
\label{eq:bounds_NDEI_0}
    \max \left\{\begin{array}{l}
-p_{0 0 \cdot 1}+p_{0 1 \cdot 0} -p_{0 1 \cdot 1} -p_{1 0 \cdot 0} -p_{1 0 \cdot 1},  \\
-2+2p_{0 0 \cdot 0}+p_{0 1 \cdot 0} +p_{1 0 \cdot 0}+p_{1 0 \cdot 1}, \\
-1+p_{0 0 \cdot 0}-p_{0 1 \cdot 0} 
\end{array}\right\} &\leq SDE(0) \\
    &\leq \min \left\{\begin{array}{l}
2p_{0 0 \cdot 0}+p_{0 1 \cdot 0} -p_{0 1 \cdot 1} +p_{1 0 \cdot 0},  \\
p_{0 0 \cdot 0}-p_{0 1 \cdot 0},  \\
1-p_{0 0 \cdot 1}+p_{0 1 \cdot 0} -p_{1 0 \cdot 0} 
\end{array}\right\}.
\end{split}
\end{align}

These bounds are again identical to those for the NDE(0) as given in \citet{sjolander2009bounds}. 

Similarly, we obtain the following bounds for $SIE(1)$ under Setting I, of
\begin{align}
\begin{split}
\label{eq:bounds_NIEI_1}
   \max \left\{\begin{array}{l}
-p_{0 0 \cdot 1} -p_{0 1 \cdot 1},  \\
-p_{0 0 \cdot 0}-p_{0 0 \cdot 1} -p_{1 0 \cdot 0}, \\
-1 + p_{0 0 \cdot 0}-p_{0 1 \cdot 1} +p_{1 0 \cdot 0}
\end{array}\right\} & \leq SIE(1) \\
   &\leq \min \left\{\begin{array}{l}
2 - p_{0 0 \cdot 0}-p_{0 0 \cdot 1} -p_{0 1 \cdot 1} -p_{1 0 \cdot 0}-p_{1 0 \cdot 1},   \\
1-p_{0 0 \cdot 1}-p_{0 1 \cdot 1},  \\
p_{0 0 \cdot 0}+p_{1 0 \cdot 0} +p_{1 0 \cdot 1} 
\end{array}\right\},
\end{split}
\end{align}

and the following bounds for $SIE(0)$ under Setting I, of
\begin{align}
\begin{split}
\label{eq:bounds_NIEI_0}
    \max \left\{\begin{array}{l}
-p_{0 0 \cdot 1}-p_{1 0 \cdot 0} -p_{1 0 \cdot 1},  \\
-2+2p_{0 0 \cdot 0}+p_{0 0 \cdot 1} +p_{0 1 \cdot 0}+p_{1 0 \cdot 0}+ p_{1 0 \cdot 1}, \\
-1+p_{0 0 \cdot 0}+p_{0 1 \cdot 0} 
\end{array}\right\} &\leq SIE(0) \\
    &\leq \min \left\{\begin{array}{l}
1-p_{0 0 \cdot 1}+p_{0 1 \cdot 0} -p_{1 0 \cdot 1},  \\
p_{0 0 \cdot 0}+p_{0 1 \cdot 0},  \\
p_{0 0 \cdot 0}+p_{0 0 \cdot 1} +p_{1 0 \cdot 1} 
\end{array}\right\}.
\end{split}
\end{align}

\subsection{Setting II}

Let $p_{yml\cdot a}=Pr(Y=y, M=m, L=l \mid A=a)$. Under (a)-(c), we obtain the following bounds for SDE$(1)$, 

\begin{align}
\begin{split}
\label{eq:bounds_SDE_II_1}
  \mbox{SDE}(1) &\geq \max \left\{\begin{array}{l}
-2 p_{0 0 0 \cdot 1}-2 p_{0 1 0 \cdot 1}- p_{1 0 0 \cdot 1}- p_{1 1 0 \cdot 1}-2 p_{0 0 1 \cdot 1}+p_{0 1 1 \cdot 0} -p_{0 1 1 \cdot 1} -p_{1 0 1\cdot 1},  \\
-1-p_{0 0 0\cdot 1}-p_{0 1 0\cdot 1} +p_{0 0 1 \cdot 0} -p_{0 1 1\cdot 1}+p_{1 0 1\cdot 1},   \\
-1-p_{0 0 0\cdot 1}+p_{0 1 0\cdot 0} +p_{1 1 0 \cdot 1} -p_{0 0 1\cdot 1}-p_{0 1 1\cdot 1}, \\ 
-1+p_{0 0 0\cdot 0}-p_{0 1 0\cdot 1} +p_{1 0 0 \cdot 1} -p_{0 0 1\cdot 1}-p_{0 1 1\cdot 1}, \\ 
-p_{0 0 0\cdot 1}-p_{0 1 0\cdot 1} -p_{0 0 1 \cdot 1} -p_{0 1 1\cdot 1} \\ 
\end{array}\right\}  \\
  \mbox{SDE}(1) &\leq \min \left\{\begin{array}{l}
2 - p_{0 0 0 \cdot 1} -2p_{0 1 0 \cdot 1} -p_{1 1 0 \cdot 0} - p_{1 1 0 \cdot 1}-p_{0 0 1 \cdot 1} -p_{0 1 1 \cdot 1},   \\
2 - p_{0 0 0 \cdot 1} -p_{0 1 0 \cdot 1} -2p_{0 0 1 \cdot 1} - p_{0 1 1 \cdot 1}-p_{10 1 \cdot 0} -p_{1 0 1 \cdot 1},   \\
p_{0 0 0 \cdot 0} + p_{0 1 0 \cdot 0} +p_{1 0 0 \cdot 0} + p_{1 0 0 \cdot 1}+p_{1 1 0 \cdot 0} +p_{1 1 0 \cdot 1}  +p_{0 0 1 \cdot 0}  +p_{0 1 1 \cdot 0}  - \\p_{0 1 1 \cdot 1}   +p_{1 0 1 \cdot 0}  +p_{1 0 1 \cdot 1},     \\
1 - p_{0 0 0 \cdot 1} -p_{0 1 0 \cdot 1} -p_{0 0 1 \cdot 1} - p_{0 1 1 \cdot 1},  \\
2 - 2p_{0 0 0 \cdot 1} -p_{0 1 0 \cdot 1} -p_{1 0 0 \cdot 0} - p_{1 0 0 \cdot 1}-p_{0 0 1 \cdot 1} -p_{0 1 1 \cdot 1}   
\end{array}\right\}.
\end{split}
\end{align}

These are identical to the bounds for the NDE$(1,1,1)$, for two sequential mediators given in \cite{Gabriel23}. In the same setting, we obtain the following bounds for SDE$(0)$, 

\begin{align}
\begin{split}
\label{eq:bounds_SDE_II_0}
  \mbox{SDE}(0) &\geq \max \left\{\begin{array}{l}
-p_{0 0 0 \cdot 1}-p_{0 1 0 \cdot 1}- p_{1 0 0 \cdot 0}- p_{1 0 0 \cdot 1}-p_{1 1 0 \cdot 0}-p_{1 1 0 \cdot 1}-p_{0 0 1 \cdot 1}+p_{0 1 1 \cdot 0} - \\ p_{0 1 1 \cdot 1}-p_{1 0 1\cdot 0}  -p_{1 0 1\cdot 1},  \\
-2+p_{0 0 0\cdot 0}+p_{0 1 0\cdot 0} +2p_{0 0 1 \cdot 0} +p_{0 1 1\cdot 0}+p_{1 0 1\cdot 0} +p_{1 0 1\cdot 1},   \\
-2+p_{0 0 0\cdot 0}+2p_{0 1 0\cdot 0} +p_{1 1 0 \cdot 0}+p_{1 1 0 \cdot 1} +p_{0 0 1\cdot 0}+p_{0 1 1\cdot 0}, \\ 
-2+2p_{0 0 0\cdot 0}+2_{0 1 0\cdot 0} +p_{1 0 0 \cdot 0}+p_{1 0 0 \cdot 1} +p_{0 0 1\cdot 0}+p_{0 1 1\cdot 0}, \\ 
-1+p_{0 0 0\cdot 0}+2_{0 1 0\cdot 0} +p_{0 0 1 \cdot 0}+p_{0 1 1 \cdot 0} 
\end{array}\right\}  \\
  \mbox{SDE}(0) &\leq \min \left\{\begin{array}{l}
1 + p_{0 0 0 \cdot 0} -p_{0 1 0 \cdot 1} -p_{1 1 0 \cdot 0} - p_{0 0 1 \cdot 0} +p_{0 1 1 \cdot 0},   \\
1 + p_{0 0 0 \cdot 0} +p_{0 1 0 \cdot 0} -p_{0 0 1 \cdot 1} - p_{0 1 1 \cdot 0} -p_{1 0 1 \cdot 0},   \\
2p_{0 0 0 \cdot 0} + 2p_{0 1 0 \cdot 0} +p_{1 0 0 \cdot 0} + p_{1 1 0 \cdot 0}+2p_{0 0 1 \cdot 0} +p_{0 1 1 \cdot 0}  -p_{0 1 1 \cdot 1}  +p_{1 0 1 \cdot 0},  \\
p_{0 0 0 \cdot 0}+p_{0 1 0 \cdot 0} +p_{0 0 1 \cdot 0} + p_{0 1 1 \cdot 0},    \\
1 - p_{0 0 0 \cdot 1} +p_{0 1 0 \cdot 0} -p_{1 0 0 \cdot 0} + p_{0 0 1 \cdot 0} +p_{0 1 1 \cdot 0}    
\end{array}\right\}. 
\end{split}
\end{align}

The bounds are identical to those for natural direct effect, NDE$(0,0,0)$, for two sequential mediators given in \cite{Gabriel23}.

These natural direct effects NDE$(0,0,0)$ and NDE$(1,1,1)$ are identified when $\boldsymbol{U}$ is measured in (a), as demonstrated in \citep{daniel2015causal}. However, there are many other versions of NDEs in the sequential mediator setting as outlined in \citet{daniel2015causal} and bounded in \citet{Gabriel23}; any natural direct effect  $NDE(x',x'',x''')$ such that $x'=x'''$ is nonparametrically identified under cross-world independence assumptions. 

Under (a)-(c), we obtain the following bounds for the $\mbox{SIE}(1)$
\begin{align}
\begin{split}
\label{eq:bounds_SIE_II_1}
  \mbox{SIE}(1) &\geq \max \left\{\begin{array}{l}
-p_{0 0 0 \cdot 1}- p_{0 0 1 \cdot 1}- p_{0 1 0 \cdot 1}- p_{0 1 1 \cdot 1},  \\
-p_{0 0 0\cdot 0}-p_{0 0 0\cdot 1} -p_{0 0 1 \cdot 0} -p_{0 0 1\cdot 1}-p_{0 1 0\cdot 0}-p_{0 1 0 \cdot 1} -p_{1 0 0 \cdot 0} -p_{1 0 1 \cdot 0} -p_{1 1 0 \cdot 0},   \\
-1-p_{0 0 0\cdot 1}-p_{0 0 1\cdot 1} + p_{0 1 0 \cdot 0} -p_{0 1 1\cdot 1}+p_{1 1 0\cdot 0}, \\ 
-1-p_{0 0 0\cdot 1}+p_{0 0 1\cdot 0} -p_{0 1 0 \cdot 1} -p_{0 1 1\cdot 1}+p_{1 0 1\cdot 0}, \\ 
-1 + p_{0 0 0\cdot 0}-p_{0 0 1\cdot 1} -p_{0 1 0 \cdot 1} -p_{0 1 1\cdot 1}+p_{1 0 0\cdot 0}, \\ 
\end{array}\right\}  \\
  \mbox{SIE}(1) &\leq \min \left\{\begin{array}{l}
2 - p_{0 0 0 \cdot 0} - p_{0 0 0 \cdot 1} - p_{0 0 1 \cdot 1} - p_{0 1 0 \cdot 1}-p_{0 1 1 \cdot 1} -p_{1 0 0 \cdot 0}-p_{1 0 0 \cdot 1},   \\
1 - p_{0 0 0 \cdot 1} -p_{0 0 1 \cdot 1} -p_{0 1 0 \cdot 1} - p_{0 1 1 \cdot 1},   \\
2 - p_{0 0 0 \cdot 1} - p_{0 0 1 \cdot 0} -p_{0 0 1 \cdot 1} - p_{0 1 0 \cdot 1}-p_{0 1 1 \cdot 1} -p_{1 0 1 \cdot 0}  -p_{1 0 1 \cdot 1}, \\
2 - p_{0 0 0 \cdot 1} - p_{0 0 1 \cdot 1} -p_{0 1 0 \cdot 0} - p_{0 1 0 \cdot 1}-p_{0 1 1 \cdot 1} -p_{1 1 0 \cdot 0} -p_{1 1 0 \cdot 1}, \\
p_{0 0 0 \cdot 0} + p_{0 0 1 \cdot 0} + p_{0 1 0 \cdot 0} + p_{1 0 0 \cdot 0}+p_{1 0 0 \cdot 1}+p_{1 0 1 \cdot 0}   +p_{1 0 1 \cdot 1} +p_{1 1 0 \cdot 0}  +p_{1 1 0 \cdot 1}    
\end{array}\right\},
\end{split}
\end{align}

and we obtain the following bounds for the $\mbox{SIE}(0)$

\begin{align}
\begin{split}
\label{eq:bounds_SIE_II_0}
  \mbox{SIE}(0) &\geq \max \left\{\begin{array}{l}
    -p_{000 \cdot 1} - p_{001 \cdot 1} - p_{010 \cdot 1} - p_{100 \cdot 0} - p_{100 \cdot 1} - p_{101 \cdot 0} - p_{101 \cdot 1} - p_{110 \cdot 0} - p_{110 \cdot 1}, \\
    -2 + p_{000 \cdot 0} + p_{001 \cdot 0} + p_{010 \cdot 0} + p_{010 \cdot 1} + p_{011 \cdot 0} + p_{110 \cdot 0} + p_{110 \cdot 1},\\
    -2 + p_{000 \cdot 0} + p_{001 \cdot 0} + p_{001 \cdot 1} + p_{010 \cdot 0} + p_{011 \cdot 0} + p_{101 \cdot 0} + p_{101 \cdot 1},\\
    -2 + p_{000 \cdot 0} + p_{000 \cdot 1} + p_{001 \cdot 0} + p_{010 \cdot 0} + p_{011 \cdot 0} + p_{100 \cdot 0} + p_{100 \cdot 1},\\
    -1 + p_{000 \cdot 0} + p_{001 \cdot 0} + p_{010 \cdot 0} + p_{011 \cdot 0}
\end{array}\right\}  \\
  \mbox{SIE}(0) &\leq \min \left\{\begin{array}{l}
    1 - p_{000 \cdot 1} + p_{001 \cdot 0} + p_{010 \cdot 0} + p_{011 \cdot 0} - p_{100 \cdot 1}, \\
    p_{000 \cdot 0} + p_{001 \cdot 0} + p_{010 \cdot 0} + p_{011 \cdot 0},
    1 + p_{000 \cdot 0} - p_{001 \cdot 1} + p_{010 \cdot 0} + p_{011 \cdot 0} - p_{101 \cdot 1}, \\
    1 + p_{000 \cdot 0} + p_{001 \cdot 0} - p_{010 \cdot 1} + p_{011 \cdot 0} - p_{110 \cdot 1}, \\
    p_{000 \cdot 0} + p_{000 \cdot 1} + p_{001 \cdot 0} + p_{001 \cdot 1} + p_{010 \cdot 0} + p_{010 \cdot 1} + p_{100 \cdot 1} + p_{101 \cdot 1} + p_{110 \cdot 1}  
\end{array}\right\}. 
\end{split}
\end{align}

\section{Comparison between Sjölander bounds and R\&R bounds} \label{app:NDEbounds_proof}
In this section, we will show that the bounds for the NDE given by \cite{sjolander2009bounds} are always at least as wide as the bounds given by \cite{RobinsRichardson2011}.

The Sjölander bounds can be rewritten as follows.
\small
        \begin{multline*}
        \mbox{UB}^{S} \\=\min \left\{\begin{array}{l}
        Pr(Y=0 \mid A=0), \\
        Pr(Y=0 \mid A=0) + Pr(M = 1 \mid A=1) - Pr(M=0 \mid A=0) + Pr(Y=1, M=0 \mid A=1), \\
        Pr(Y=0 \mid A=0) + Pr(M=0 \mid A=1) - Pr(M=1 \mid A=0) + Pr(Y=1, M=1 \mid A=1)
\end{array}\right\},
\end{multline*}
\normalsize
and
\begin{align*} 
        \mbox{LB}^{S} =\max \left\{\begin{array}{l}
        -Pr(Y=1 \mid A=0), \\
        -Pr(Y=1 \mid A=0) - Pr(M=0 \mid A=0) + Pr(Y=1, M=1 \mid A=1), \\
        -Pr(Y=1 \mid A=0) - Pr(M=1 \mid A=0) + Pr(Y=1, M=0 \mid A=1),
\end{array}\right\}.
\end{align*}

The Robins \& Richardson bounds can be written as 
\begin{align*} 
        \mbox{UB}^{RR} =&\min \left\{Pr(M=0 \mid A=0), Pr(Y=1 \mid M=0, A=1) \right\}\\
        &+\min \left\{Pr(M=1 \mid A=0), Pr(Y=1 \mid M=1, A=1) \right\} - Pr(Y=1 \mid A=0),
\end{align*}
and
\begin{align*} 
        \mbox{LB}^{RR} =&\max \left\{0, Pr(M=0 \mid A=0)- Pr(Y=0 \mid M=0, A=1)  \right\}\\
        &+\max \left\{0, Pr(M=1 \mid A=0)- Pr(Y=0 \mid M=1, A=1) \right\} - Pr(Y=1 \mid A=0).
\end{align*}

\subsection{Upper bound proof} 
First, we show that $\mbox{UB}^{RR} \leq \mbox{UB}^{S}$. We consider the four possible terms of $\mbox{UB}^{RR}$ separately.


\subsubsection[Setting I]{$\mbox{UB}^{RR} = Pr(Y=0 \mid A=0)$}
    We have
    \begin{align} \label{ineq1}
        &Pr(M=0 \mid A=0) < Pr(Y=1 \mid A=1, M=0) \nonumber\\ 
        &\qquad \Leftrightarrow 1- Pr(M=1 \mid A=0) < 1- Pr(Y=0 \mid A=1, M=0) \nonumber\\ 
        &\qquad \Leftrightarrow - Pr(M=1 \mid A=0) < - Pr(Y=0 \mid A=1, M=0) < - Pr(Y=0, M=0 \mid A=1)\nonumber\\ 
        &\qquad \Leftrightarrow - Pr(M=1 \mid A=0) + Pr(Y=0, M=0 \mid A=1)<0\nonumber\\ 
        &\qquad \Leftrightarrow - Pr(M=1 \mid A=0) +Pr(Y=0 \mid A=1) - Pr(Y=0, M=1 \mid A=1) < 0\nonumber\\
        &\qquad \Leftrightarrow  Pr(M=1 \mid A=0) - Pr(Y=0 \mid A=1) + Pr(Y=0, M=1 \mid A=1) > 0,
    \end{align}
    and similarly 
    \begin{multline}\label{ineq2}
                Pr(M=1 \mid A=0) < Pr(Y=1 \mid A=1, M=1) \\ 
                \Leftrightarrow  - Pr(M=0 \mid A=0) < - Pr(Y=0, M=1 \mid A=1)\\
                \Leftrightarrow  Pr(M=0 \mid A=0) - Pr(Y=0 \mid A=1) + Pr(Y=0, M=0 \mid A=1) > 0.
    \end{multline}

    We can rewrite the Sjölander upper bounds as 
    \small
        \begin{multline*} 
        \mbox{UB}^{S} \\ =\min \left\{\begin{array}{l}
        Pr(Y=0 \mid A=0), \\
        Pr(Y=0 \mid A=0) + Pr(M=1 \mid A=0) - Pr(Y=0 \mid A=1) + Pr(Y=0, M=1 \mid A=1), \\
        Pr(Y=0 \mid A=0)  + Pr(M=0 \mid A=0) - Pr(Y=0 \mid A=1) + Pr(Y=0, M=0 \mid A=1)
\end{array}\right\}.
    \end{multline*}
    \normalsize

    So by \eqref{ineq1} and \eqref{ineq2} we have $\mbox{UB}^{S} = Pr(Y=0 \mid A=0) = \mbox{UB}^{RR}$.

\subsubsection[Setting II]{$\mbox{UB}^{RR} =  Pr(Y=1 \mid M=0, A=1)+Pr(Y=1 \mid M=1, A=1) - Pr(Y=1 \mid A=0)$}
    We have
    \begin{align} \label{ineq3}
        Pr(M=0 \mid A=0) > Pr(Y=1 \mid A=1, M=0) > Pr(Y=1, M=0 \mid A=1),
    \end{align}
    and
    \begin{align} \label{ineq4}
                Pr(M=1 \mid A=0) > Pr(Y=1 \mid A=1, M=1) > Pr(Y=1, M=1 \mid A=1).
    \end{align}

    From \eqref{ineq3} and \eqref{ineq4} it follows that
    \begin{align*}
        \mbox{UB}^{RR} &=  Pr(Y=1 \mid M=0, A=1)+Pr(Y=1 \mid M=1, A=1) - Pr(Y=1 \mid A=0) \\
        &< Pr(M=0 \mid A=0)+ Pr(M=1 \mid A=0) - Pr(Y=1 \mid A=0) \\
        &= Pr(Y=0 \mid A=0),
    \end{align*}
    and
    \begin{align*}
        \mbox{UB}^{RR} &=  Pr(Y=1 \mid M=0, A=1)+Pr(Y=1 \mid M=1, A=1) - Pr(Y=1 \mid A=0) \\
        &=Pr(Y=0 \mid A=0) - Pr(Y=0 \mid M=0, A=1)+Pr(Y=1 \mid M=1, A=1) \\
        &< Pr(Y= 0 \mid A=0) - Pr(Y=0, M=0 \mid A=1) + Pr(M=1 \mid A=0) \\
        &= Pr(Y=0 \mid A=0) + Pr(M=1 \mid A=1) - Pr(M=0 \mid A=0) + Pr(Y=1, M=0 \mid A=1),
    \end{align*}
    and
        \begin{align*}
        \mbox{UB}^{RR} &=  Pr(Y=1 \mid M=0, A=1)+Pr(Y=1 \mid M=1, A=1) - Pr(Y=1 \mid A=0) \\
        &=Pr(Y=0 \mid A=0) + Pr(Y=1 \mid M=0, A=1)-Pr(Y=0 \mid M=1, A=1)\\
        &<Pr(Y=0 \mid A=0) + Pr(M=0 \mid A=0)-Pr(Y=0, M=1 \mid A=1) \\
        &=Pr(Y=0 \mid A=0) - Pr(M=1 \mid A=0)+Pr(M=0 \mid A=1)+ Pr(Y=1, M=1 \mid A=1).
    \end{align*}

    So, $\mbox{UB}^{RR}$ is strictly smaller than all three terms in the Sjölander upper bound.

    \subsubsection[Setting III]{$\mbox{UB}^{RR} =  Pr(Y=0 \mid A=0) - Pr(M=1 \mid A=0) + Pr(Y=1 \mid A=1, M=1)$}
    We have as in \eqref{ineq1}
    \begin{multline} \label{ineq5}
        Pr(M=0 \mid A=0) < Pr(Y=1 \mid A=1, M=0) \\ 
        \Leftrightarrow  Pr(M=1 \mid A=0) - Pr(Y=0 \mid A=1) + Pr(Y=0, M=1 \mid A=1) > 0.
    \end{multline}
    We also have as in \eqref{ineq4}
    \begin{multline} \label{ineq6}
        Pr(M=1 \mid A=0) > Pr(Y=1 \mid A=1, M=1)  \\
                 \Leftrightarrow  Pr(Y=1, M=1 \mid A=1) - Pr(M=1 \mid A=0) < 0.
    \end{multline}

        We can rewrite the Sjölander bounds as 
        \small
        \begin{multline*}
        \mbox{UB}^{S} \\=\min \left\{\begin{array}{l}
        Pr(Y=0 \mid A=0), \\
        Pr(Y=0 \mid A=0) + Pr(M=1 \mid A=0) - Pr(Y=0 \mid A=1) + Pr(Y=0, M=1 \mid A=1), \\
               Pr(Y=0 \mid A=0) - Pr(M=1 \mid A=1) + Pr(M=0 \mid A=0) + Pr(Y=1, M=1 \mid A=1)
\end{array}\right\},
\end{multline*}
\normalsize

From \eqref{ineq5} the second term is larger than the first term and cannot be the minimum. From \eqref{ineq6} we have 
\begin{align*}
    \mbox{UB}^{RR} =  Pr(Y=0 \mid A=0) - Pr(M=1 \mid A=0) + Pr(Y=1 \mid A=1, M=1) <Pr(Y=0 \mid A=0) ,
\end{align*}
so $\mbox{UB}^{RR}$ is smaller than the first term in $\mbox{UB}^{S}$.
We also have
\begin{align*}
    \mbox{UB}^{RR} &=  Pr(Y=0 \mid A=0) - Pr(M=1 \mid A=0) + Pr(Y=1 \mid A=1, M=1) \\
    &=  Pr(Y=0 \mid A=0) + Pr(M=0 \mid A=0) - Pr(Y=0 \mid A=1, M=1) \\
    &<  Pr(Y=0 \mid A=0) + Pr(M=0 \mid A=0) - Pr(Y=0, M=1 \mid A=1) \\
    &=  Pr(Y=0 \mid A=0) + Pr(M=0 \mid A=0) - Pr(M=1 \mid A=1) + Pr(Y=1, M=1 \mid A=1) 
\end{align*}
so $\mbox{UB}^{RR}$ is smaller than the third term in $\mbox{UB}^{S}$

    \subsubsection[Setting IV]{$\mbox{UB}^{RR} =  Pr(Y=0 \mid A=0) - Pr(M=0 \mid A=0) + Pr(Y=1 \mid A=1, M=0)$}
    This is symmetric to the scenario above. 

\subsection{Lower bound proof}
Next, we show that $\mbox{LB}^{RR} \geq \mbox{LB}^{S}$. We consider the four possible terms of $\mbox{LB}^{RR}$ separately.


    \subsubsection[Setting I]{$\mbox{LB}^{RR}=-Pr(Y=1 \mid A=0)$}
    When $\mbox{LB}^{RR}=-Pr(Y=1 \mid A=0)$ we have 
    \begin{multline}\label{ineq7}
        Pr(M=0 \mid A=0) < Pr(Y=0 \mid M=0, A=1) \\
        \Leftrightarrow -Pr(M=1 \mid A=0) + Pr(Y=1, M=0 \mid A=1) <0, 
    \end{multline}
    and
        \begin{multline}\label{ineq8}
        Pr(M=1 \mid A=0) < Pr(Y=0 \mid M=1, A=1) \\
        \Leftrightarrow -Pr(M=0 \mid A=0) + Pr(Y=1, M=1 \mid A=1) <0, 
    \end{multline}

    From \eqref{ineq7} and \eqref{ineq8} we have $\mbox{LB}^S = -Pr(Y=1 \mid A=0) = \mbox{LB}^{RR}$

    \subsubsection[Setting II]{ $\mbox{LB}^{RR}=1-Pr(Y=0 \mid M=0, A=1)-Pr(Y=0 \mid M=1, A=1)-Pr(Y=1 \mid A=0)$}
    We have
    \begin{align}\label{ineq9}
        Pr(M=0 \mid A=0) > Pr(Y=0 \mid M=0, A=1) > Pr(Y=0,  M=0 \mid A=1) ,
    \end{align}
    and
        \begin{align}\label{ineq10}
        Pr(M=1 \mid A=0) > Pr(Y=0 \mid M=1, A=1) >Pr(Y=0, M=1 \mid  A=1).
    \end{align}

    From \eqref{ineq9} and \eqref{ineq10} it follows that
    \begin{align*}
        \mbox{LB}^{RR}&=1-Pr(Y=0 \mid M=0, A=1)-Pr(Y=0 \mid M=1, A=1)-Pr(Y=1 \mid A=0) \\
        &> 1- Pr(M=0 \mid A=0) - Pr(M=1 \mid A=1)  -Pr(Y=1 \mid A=0)  \\
        &=-Pr(Y=1 \mid A=0),
    \end{align*}
    and
    \begin{align*}
        \mbox{LB}^{RR}&=1-Pr(Y=0 \mid M=0, A=1)-Pr(Y=0 \mid M=1, A=1)-Pr(Y=1 \mid A=0) \\
        &=-Pr(Y=1 \mid A=0)  + Pr(Y=1 \mid M=0, A=1)-Pr(Y=0 \mid M=1, A=1)\\
        &> -Pr(Y=1 \mid A=0)  +Pr(Y=1, M=0 \mid A=1) - Pr(M=1 \mid A=0), 
    \end{align*}
    and
        \begin{align*}
        \mbox{LB}^{RR}&=1-Pr(Y=0 \mid M=0, A=1)-Pr(Y=0 \mid M=1, A=1)-Pr(Y=1 \mid A=0) \\
        &=-Pr(Y=1 \mid A=0) -Pr(Y=0 \mid M=0, A=1) +Pr(Y=1 \mid M=1, A=1)\\
        &> -Pr(Y=1 \mid A=0) - Pr(M=0 \mid A=0) + Pr(Y=1, M=1 \mid A=1).
    \end{align*}
        so $\mbox{LB}^{RR}$ is greater than all three terms in the lower bound of Sjölander. 

    \subsubsection[Setting III]{$\mbox{LB}^{RR}=Pr(M=1 \mid A=0)-Pr(Y=0 \mid M=1, A=1)-Pr(Y=1 \mid A=0)$}  
    We have as in \eqref{ineq7}
        \begin{multline}\label{ineq11}
        Pr(M=0 \mid A=0) < Pr(Y=0 \mid M=0, A=1) \\
        \Leftrightarrow -Pr(M=1 \mid A=0) + Pr(Y=1, M=0 \mid A=1) <0, 
    \end{multline}
    and as in \eqref{ineq10}
        \begin{align}\label{ineq12}
        Pr(M=1 \mid A=0) > Pr(Y=0 \mid M=1, A=1) >Pr(Y=0, M=1 \mid  A=1).
    \end{align}

    From \eqref{ineq11} the third term in $\mbox{LB}^S$ is smaller than the first term and cannot be the maximum. From \eqref{ineq12} we have
    \begin{align*}
        \mbox{LB}^{RR}&=Pr(M=1 \mid A=0)-Pr(Y=0 \mid M=1, A=1)-Pr(Y=1 \mid A=0) \\
        &>-Pr(Y=1 \mid A=0),
    \end{align*}
    so $\mbox{LB}^{RR}$ is larger than the first term in $\mbox{LB}^S$. We also have
    \begin{align*}
        \mbox{LB}^{RR}&=Pr(M=1 \mid A=0)-Pr(Y=0 \mid M=1, A=1)-Pr(Y=1 \mid A=0) \\
        &=-Pr(M=0 \mid A=0)+Pr(Y=1 \mid M=1, A=1)-Pr(Y=1 \mid A=0)\\
        &>-Pr(M=0 \mid A=0)+Pr(Y=1, M=1 \mid A=1)-Pr(Y=1 \mid A=0),
    \end{align*}
    so $\mbox{LB}^{RR}$ is larger than the second term in $\mbox{LB}^S$.

    \subsubsection[Setting IV]{$\mbox{LB}^{RR}=Pr(M=0 \mid A=0)-Pr(Y=0 \mid M=0, A=1)-Pr(Y=1 \mid A=0)$}
    This is symmetric to the scenario above.

\section{Numeric comparison of bounds under Setting II} \label{app:NDEnumeric}

\begin{figure}
    \centering
    \includegraphics[width=0.55\linewidth]{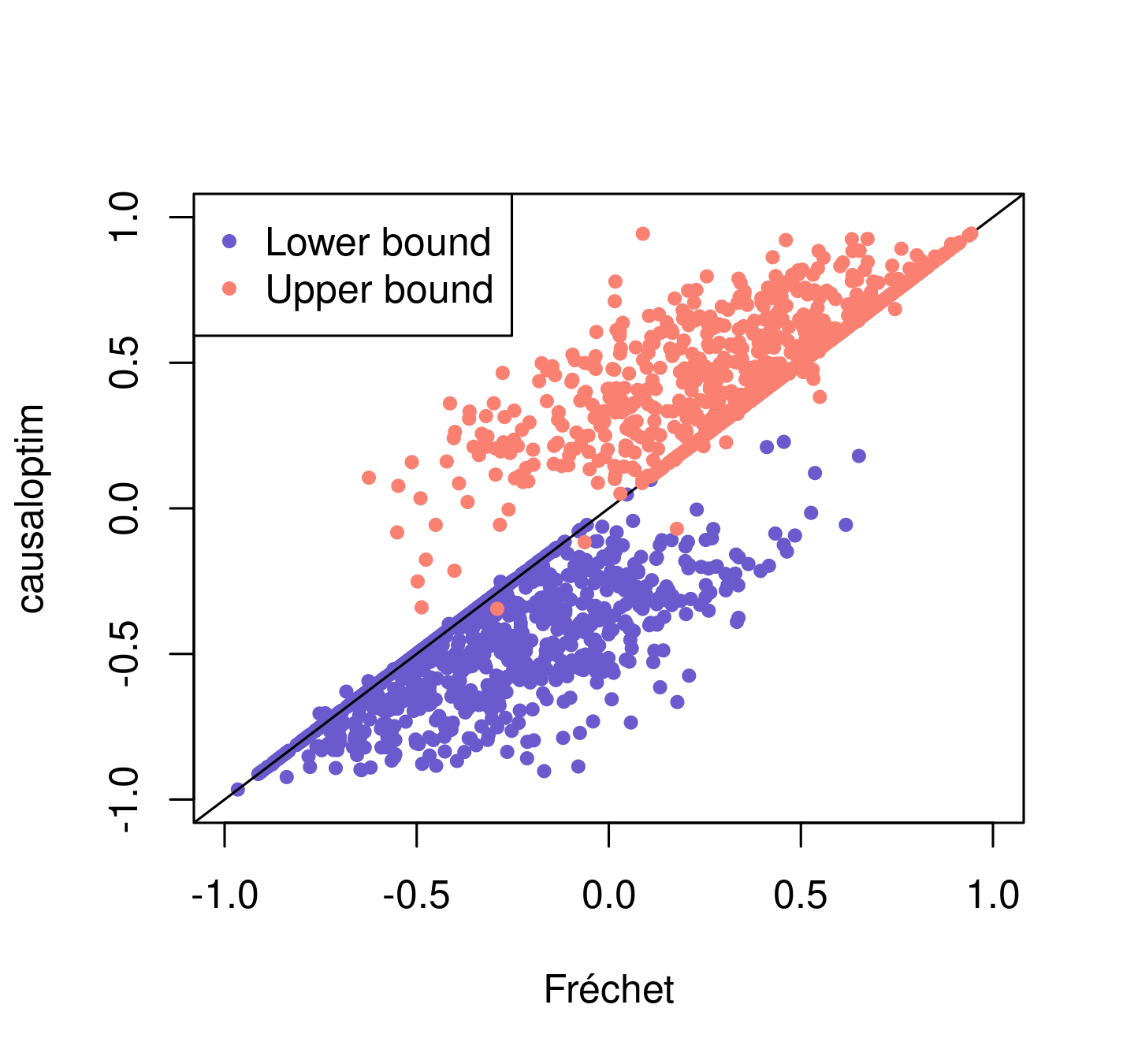}
    \caption{Comparison of Fréchet versus causaloptim bounds in Setting II for the NDE(0,0,0). Each point represents a lower or upper bound calculated under one of 1000 random distributions where $Pr(Y = 1| A, L, M), Pr(M = 1| A, L), $and $Pr(L = 1| A)$ are all uniform on (0, 1). Such distributions are compatible with the sequential mediation setting where there is no unmeasured confounder. }
    \label{fig:ndenumeric}
\end{figure}

\end{document}